%% file: ieee.tex
\documentclass[10pt,conference]{IEEEtran}
\IEEEoverridecommandlockouts

\usepackage{graphicx}
\graphicspath{Images/}

\usepackage{cite}
\usepackage{xcolor}
\def\BibTeX{{\rm B\kern-.05em{\sc i\kern-.025em b}\kern-.08em
    T\kern-.1667em\lower.7ex\hbox{E}\kern-.125emX}}

\usepackage{amsmath,amsfonts,amssymb,amsthm}
\usepackage{mathtools}
\usepackage{booktabs}
\usepackage{float}
\usepackage[normalem]{ulem}
\usepackage[ruled,vlined]{algorithm2e}
\usepackage[nocomma]{optidef}
\usepackage[normalem]{ulem}

\usepackage{enumitem}

\usepackage{caption}
\usepackage{subcaption}
\usepackage{subfiles}

\usepackage[outdir=./]{epstopdf}

\def\BibTeX{{\rm B\kern-.05em{\sc i\kern-.025em b}\kern-.08em
    T\kern-.1667em\lower.7ex\hbox{E}\kern-.125emX}}

\SetCommentSty{mycommfont}

\DeclareMathOperator{\E}{\mathbb{E}}
\DeclareMathOperator{\Var}{\textrm{Var}}
\DeclareMathOperator{\Cov}{\textrm{Cov}}
\newcommand{\R}{\mathbb{R}}
\newcommand{\nr}{n_{r}}
\newcommand{\ns}{n_{s}}
\newcommand{\nri}{n_{r,i}}
\newcommand{\nsi}{n_{s,i}}
\newcommand{\npi}{n_{p,i}}
\newcommand{\probl}{\textrm{equation 1}}
\newcommand{\probs}{\textrm{eq. 1}}

\makeatletter
\newcommand{\customlabel}[2]{%
\protected@write \@auxout {}{\string \newlabel {#1}{{#2}{}}}}
\makeatother

\newcommand{\del}[1]{}

\renewcommand{\footnoterule}{%
  \kern -3pt
  \hrule width \columnwidth height 1pt
  \kern 2pt
}

\AtBeginDocument{%
  \providecommand\BibTeX{{%
    \normalfont B\kern-0.5em{\scshape i\kern-0.25em b}\kern-0.8em\TeX}}}

\begin{document}

\title{Efficient Transmission and Reconstruction of Dependent Data Streams via Edge Sampling}

\author{\IEEEauthorblockN{Joel Wolfrath and Abhishek Chandra}
\IEEEauthorblockA{Department of Computer Science and Engineering\\
University of Minnesota,
Minneapolis, USA\\
Email: \{wolfr046, chandra\}@umn.edu\\}
}

\maketitle
\thispagestyle{plain}
\pagestyle{plain}

\begin{abstract}
\subfile{Sections/abstract}
\end{abstract}

\begin{IEEEkeywords}
Stream processing, edge computing, big data, approximate computing
\end{IEEEkeywords}

\section{Introduction}
\label{sec:intro}

\subfile{Sections/intro}

\section{Preliminaries}
\label{sec:prelim}

\subfile{Sections/prelim}

\section{Proposed Approach}
\label{sec:approach}

\subfile{Sections/approach}

\section{Practical Heuristics}
\label{sec:heuristic}

\subfile{Sections/heuristics}

\section{Evaluation}
\label{sec:eval}

\subfile{Sections/eval}

\section{Related Work}
\label{sec:related}

\subfile{Sections/related}

\section{Conclusion}
\label{sec:conclusion}

\subfile{Sections/conclusion}

\bibliographystyle{IEEEtran}
\bibliography{ieee}

\appendices

\subfile{Sections/appendix}

\end{document}

%% file: Sections/abstract.tex
Data stream processing is an increasingly important topic due to the prevalence of smart devices and the demand for real-time analytics.
Geo-distributed streaming systems, where cloud-based queries utilize data streams from multiple distributed devices, face challenges since wide-area network (WAN) bandwidth is often scarce or expensive.
Edge computing allows us to address these bandwidth costs by utilizing resources close to the devices, e.g. to perform sampling over the incoming data streams, which trades downstream query accuracy to reduce the overall transmission cost.
In this paper, we leverage the fact that correlations between data streams may exist across devices located in the same geographical region.
Using this insight, we develop a hybrid edge-cloud system which systematically trades off between sampling at the edge and estimation of missing values in the cloud to reduce traffic over the WAN.
We present an optimization framework which computes sample sizes at the edge and systematically bounds the number of samples we can estimate in the cloud given the strength of the correlation between streams.
Our evaluation with three real-world datasets shows that compared to existing sampling techniques, our system could provide comparable error rates over multiple aggregate queries while reducing WAN traffic by 27-42\%.

%% file: Sections/intro.tex
Real-time analytics and data-driven insights continue to play a pivotal role in online applications and business operations.
Organizations require efficient query mechanisms to handle increasingly large datasets and the ubiquity of smart devices.
Data streaming applications in particular continue to grow at an unprecedented rate, with projections suggesting there will be more than 75 billion connected devices by the year 2025~\cite{IHS}.
This includes personal devices, smart home components, smart cities, retail environments and industrial settings.
This enormous growth in data generated at the edge has driven much research into efficient data transfer, given the scarcity and expense associated with transferring data over the wide-area network (WAN) \cite{dhruv2,feather,ic2e}.
Recent strides in edge computing provide a mechanism for addressing
these constraints by leveraging computation close to the data-generating devices, allowing us to make optimizations and reduce cost.
While we focus on minimizing network traffic, edge resources can address many other objectives, including maximizing throughput or minimizing end-to-end latency.

Approximation techniques are often used to trade downstream query accuracy for a reduction in traffic over the WAN. Sampling at the network edge is commonly used to reduce transmissions without introducing substantial error in downstream queries~\cite{approxiot,edge_sample2, adam}.
Sampling also allows us to bound the amount of error introduced in many kinds of queries\textemdash an important property for quantifying worst-case performance in production deployments. When data arrives in the cloud, additional samples may be imputed or estimated using a model~\cite{sim,sim2}. These techniques can be effective in practice, but existing sampling algorithms focus on the properties of individual streams and models fail to provide performance guarantees when making inferences out-of-sample.

In this work, we propose a novel sampling algorithm for addressing WAN scarcity which exploits similarities in data streams collected by multiple localized devices in a geo-distributed environment. Prior research suggests that devices located in the same geographical region may produce data streams that are correlated or exhibit some kind of dependency~\cite{corr1, corr2,skewscout}. These correlations arise naturally if the sensors\footnote{We use the terms "device" and "sensor" interchangeably in this work.} are recording the same phenomenon in a geographic region (e.g. temperature), or if there is a correlation between different sensors due to external factors (e.g. human occupancy patterns in a building or neighborhood). We identify and leverage these dependencies in real time to improve sampling and modeling decisions. Identifying these dependencies at the edge allows us to explicitly control imputation accuracy, since we have access to all of the data at the edge prior to sampling. We make the following research contributions: 

\begin{itemize}
    \item We develop a hybrid edge-cloud streaming system which addresses WAN scarcity by leveraging dependencies between multiple streams in real time (without any offline profiling).
    \item We present a novel sampling algorithm and associated optimization problem for computing the number of samples to send over the network and impute in the cloud, given the dependence structure. We also provide explicit bounds on the error introduced without making any strong assumptions about the underlying data distribution.
    \item We evaluate our system across a variety of real-world datasets using a Storm-based implementation\footnote{https://github.com/jswolfrath/edge-approx.} and examine its efficacy and its sensitivity to a variety of tuning parameters.
\end{itemize}

%% file: Sections/prelim.tex
To motivate our problem, we consider applications that are inherently latency sensitive and potentially bandwidth constrained. For example, wind turbines and smart trains can be equipped with IoT devices that generate data and send it to a centralized location \cite{trains1, trains2}. This data is used to identify failures in real time, resulting in timely repairs and substantial cost savings. Furthermore, monitoring the performance output of wind turbines is important for estimating revenue in real time \cite{turbine1}. Since trains and wind turbines are often in rural areas, the available bandwidth will be scarce.
Urban settings can also benefit from bandwidth reduction techniques. Smart cities can have a very large number of sensors generating data simultaneously \cite{smart_city1,ic2e2}, which increases the traffic over the WAN. These applications highlight the importance of efficient bandwidth usage in both rural and urban settings.

\subsection{Problem Statement}

\begin{figure}
  \centering
  \includegraphics[width=0.70\columnwidth,scale=0.50,]{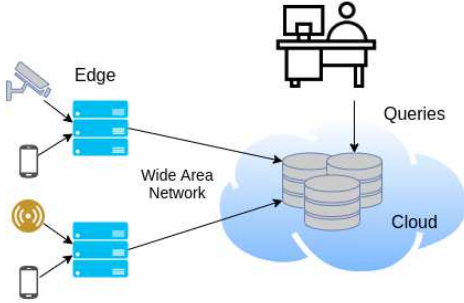}
  \caption{A geo-distributed streaming system consisting of four data-generating devices. Data is sent to two local edge nodes prior to being transferred over the WAN and persisted in the cloud.}
  \label{topology-fig}
\end{figure}

We consider a three-tiered distributed streaming framework consisting of data-generating devices, edge nodes for local processing of this data, and a destination cloud data center where the data is finally persisted. In this work, we broadly define edge nodes to include routers, base stations, or dedicated edge servers. Figure \ref{topology-fig} shows an example of this system topology. Streaming data typically consists of tuples and an associated timestamp. The key property is that the stream is unbounded, which implies the data observed at the edge is \textit{transient} in nature. In this work, we assume a tumbling window model, where data is aggregated and processed as mini-batches across fixed periods of time.
Selecting a windowing method and aggregation duration is an active area of research which heavily depends on the application and its tolerance for delay~\cite{window_spec, dhruv}.

When streaming data is persisted in the cloud, it can be represented in a variety of ways, ranging from storing simple point estimates (e.g. means or sums) to total reconstructions of the original streams.
In this work, \textit{our objective is to persist samples that minimize error when computing multiple aggregate functions}, such as counts, averages, standard deviations, and order statistics (e.g. minimum, maximum, or median). These aggregates allow us to preserve properties of the underlying data distribution without requiring a full reconstruction of the streams. The error minimization is subject to a budget constraint which dictates the amount of data we are allowed to send over the WAN in each window. Alternatively, the problem can be viewed as attempting to minimize the data transfer cost, subject to a specified tolerance for error.

\subsection{Stream Sampling}
\label{text:sampling}

In general, streaming high-frequency data over the WAN may be infeasible due to scarcity or high cost. This motivates sampling schemes at the edge which send a subset of the data while providing a bound on the error associated with downstream queries.
To ensure a fixed transmission cost, one could select a simple random sample (SRS) of the data points observed in a tumbling window and forward those points to the cloud. However, in practice, an inbound data stream may be composed of several heterogeneous \textit{sub-streams}, each with their own statistical properties \cite{heterogeneous1}. This heterogeneity justifies the use of more sophisticated sampling techniques at the edge.

When each stream has distinct statistical properties or observed frequencies, \textit{stratified sampling} may be preferable to a SRS \cite{heterogeneous1,heterogeneous2}.
Widely used in practice \cite{approxiot,approxjoin}, this technique has the desirable property of ensuring each stratum of a partitioned dataset is represented in the sample.
Some systems use stratification methods that leverage properties of the underlying data. For example, the S-VOILA system stratifies samples by assigning more samples to streams with a high degree of variability and fewer samples to streams with less variability~\cite{strat_stream}.
These stream sampling techniques provide a mechanism for streaming applications to trade downstream query accuracy for a reduction in the amount of data sent over the network.

\subsection{Imputation Strategies}
One method for handling missing values in a dataset is to \textit{impute} them, i.e., replace them with some kind of point estimate. It is common to replace missing values with a statistic (e.g. the mean) or estimate them using a model (e.g. linear regression) which is constructed using the available data. When sampling is performed at the edge, the cloud will necessarily receive a partial dataset which does not contain samples for every point in time. We propose a careful use of these imputation techniques to estimate a subset of the missing values while controlling for error. We leverage the fact that statistics or models used for imputation can be estimated at the edge, where more data is available as compared to the cloud, which can only work with the down-sampled data.

%% file: Sections/approach.tex
\subsection{Overview}

\begin{figure}
    \centering
    \includegraphics[width=0.99\columnwidth,scale=0.99]{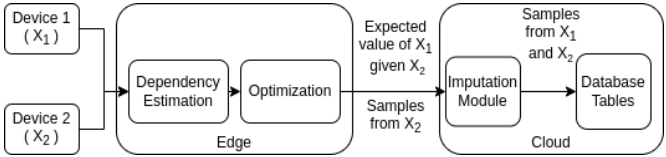}
    \caption{Proposed edge-cloud streaming framework.} 
    \label{fig-sys}
\end{figure}

We propose an edge-cloud framework for data streaming that combines edge-side sampling with cloud-based imputation of missing data values. 
We leverage edge resources to sample multiple incoming data streams and estimate dependencies between them. The edge also performs optimization to determine which samples and models to send over the network.
The cloud uses these samples and models to impute missing data values.
This framework uses the insight that streams produced by individual devices in the same geographical region may be correlated \cite{corr1,corr2,skewscout}.
For example, this is expected when devices in the same sensor network generate streaming data \cite{corr3,bbq}. Mazhar \textit{et al.} show that smart-home IoT traffic is often directly correlated with human activity patterns \cite{corr2}. Even devices in a smart city that measure different quantities could produce correlated streams \cite{corr1, corr3}. The real-world datasets we use for evaluation also substantiate this claim (see section~\ref{sec:eval} for details).
Identifying these dependencies provides an opportunity for exploiting them for efficient data transfer. 

To illustrate the opportunity, consider two data generating devices, $D_1$ and $D_2$, which are located in the same geographical region and forward their data to a single edge node. The local edge node caches the inbound streaming data for the duration of the tumbling window and then selects a subset of the received data to be forwarded to the cloud. Existing sampling techniques follow this same approach and allow us to bound the amount of error introduced for aggregation queries in each window \cite{approxiot}.

Now, suppose that the data streams generated by $D_1$ and $D_2$, represented by random variables $X_1$ and $X_2$, exhibit strong linear correlation over time. This is expected, for example, if the devices are in close proximity and capturing temperature data. We can leverage this dependency by generating a compact representation of $\E[X_1 | X_2]$ (or $X_2$ given $X_1$) at the edge and instruct the cloud to impute values based on this model. If the correlation is sufficiently strong, the edge may choose to exclusively forward samples from $D_2$ along with a compact representation of $\E[X_1 | X_2]$. The cloud can then use the model to generate samples from $D_1$ without incurring any cost for transferring them (see Figure~\ref{fig-sys}). This technique has the potential to reduce the network traffic required to obtain error rates comparable to other sampling mechanisms, provided we can determine how many samples to send and impute each window.

\subsection{Optimization Problem}
\label{sec:assumptions}

Our optimization framework allocates real samples to device streams (to forward to the cloud) and computes the number of samples to impute for each device to minimize the error introduced for aggregate queries. Table \ref{tab:Notation} outlines the notation used throughout this section with respect to a single tumbling window. We assume that the streams produce real-valued data; however, we make no strong assumptions about the data distribution of each device.
We also assume that samples drawn from each device are independent and identically distributed within a single stream window (we relax this assumption in section \ref{sec:heuristic}). Then for each stream $i$, we seek the optimal number of real and imputed samples ($\nri$ and $\nsi$), in addition to identifying a highly correlated stream to use as a predictor ($p_i$). We can then build a model for the expected value of $X_i$ given $X_{p_i}$, which allows us to impute data from stream $i$ given observations from stream $p_i$. We also include an upper bound on the number of values we should impute, given the quality of the model. Our full optimization problem is outlined in \probl. We motivate and provide intuition for the objective and constraints in the following sections.

\begin{table}[htbp]\caption{Notation}
\footnotesize
\centering 
\begin{tabular}{|p{1.0cm}|p{6.5cm}|}
 \hline
$k$ & Number of streams available for sampling \\
 \hline
$X_i$ & A random variable following the probability distribution from stream $i$ \\
 \hline
$N_i$  & The number of tuples that arrive from stream $i$ \\
 \hline
$\nri$  & Number of real samples allocated to stream $i$  \\
 \hline
$\nsi$  & Number of imputed samples allocated to stream $i$ \\
 \hline
$p_i$  & An index associated with stream $i$ corresponding to a predictor stream which correlates with stream $i$ \\
 \hline
$\mu_i$  & The mean of stream $i$ \\
 \hline
$\sigma^2_i$ & The variance of stream $i$  \\
 \hline
$w_i$ & Weights identifying a stream's importance \\
 \hline
$c_i(n,m)$ & Cost to forward $n$ samples and impute $m$ samples for stream $i$ (must be a \textit{convex} function of $n$,$m$) \\
 \hline
$C$ & Upper bound on the traffic cost \\
 \hline
$\epsilon_i$ & Bound on the reduction in variance introduced by imputation for stream $i$ \\
 \hline
\end{tabular}
\label{tab:Notation}
\end{table}

\begin{mini!}|s|[2]
    {\nr, \; \ns, \; p}
    {\sum_{i=1}^{k} w_i^2 \Var[\hat{\mu}_i]}
    {}
    {}
    \addConstraint{p_i}{\in \{1, 2, ... , k\} - \{i\} \label{eq:con1}}
    \addConstraint{0}{\leq \; \nri \; \leq \; N_i \label{eq:con2}}
    \addConstraint{0}{\leq \; \nsi \; \leq \; \npi \label{eq:con3}}
    \addConstraint{(\nri + \nsi)}{\; > 1 \label{eq:con4}}
    \addConstraint{\sum_{i=1}^k c_i(\nri,\; \nsi) }{\; \leq \; C \label{eq:con5}}
    \addConstraint{|\;  \textrm{Bias}(\hat{\sigma}_i^2) \; |}{\; \leq \; \epsilon_i \label{eq:con6}}
\end{mini!}

\subsubsection{Objective Function}
We seek to minimize error when estimating aggregate queries, subject to a bound on the traffic cost. Our problem statement requires us to minimize the error associated with many aggregate functions. However, the sampling distributions for certain aggregates (e.g. order statistics) depend on the data distribution itself, which we do not assume is known. Our proposal is to directly minimize the error associated with an AVG query while simultaneously controlling for the amount of bias introduced in our estimates for other aggregates. Minimizing the error for the mean necessarily reduces the error for many other aggregates. Errors for order statistics and percentiles are inversely proportional to the number of samples, so \textit{increasing the effective sample size through accurate imputation necessarily reduces errors for these queries} \cite{order_stats}. Therefore, minimizing a weighted\footnote{In our implementation, we set the weights to be inversely proportional to the expected value of each stream, thereby allowing us to minimize the coefficient of variation. This is a common approach which prevents us from disproportionately assigning samples to streams with high variance if the variance relative to the mean is low~\cite{sec_recommend}.} sum of the squared errors for an AVG query across all streams results in the objective:

\setcounter{equation}{1}
\begin{equation}
    f(n) = \sum_{i=1}^k w_i^2 \Var[\hat{\mu}_i] = \sum_{i=1}^k \frac{w_i^2 \sigma_{i}^2}{\nri + \nsi}
\end{equation}
\noindent
Note that this formulation does not require us to make any strong assumptions about the distribution of each stream.

\subsubsection{Constraints}

For each stream $i$, the variable $p_i$ is assigned the index of a different stream, which will be used as a candidate for prediction (constraint \ref{eq:con1}). We require that each stream have an associated predictor stream, i.e. no predictor can be null. If the predictor stream correlates poorly with the target stream, constraint \ref{eq:con6} will prevent us from performing imputation.
This formulation only allows for a single stream to be used as a predictor; we discuss the possibility of using multiple predictors in section \ref{sec:discussion}.

Constraints \ref{eq:con2} and \ref{eq:con3} provide upper bounds on the number of real and imputed samples. We can't send more real samples than we observed at the edge and we can't impute more than the number of real samples taken from the predictor stream. 

Constraint \ref{eq:con4} ensures that we have at least one sample from each stream, regardless of if it is real or imputed. This is required for constraint \ref{eq:con6} to be defined (see equation \ref{eq:bias}).

\del{Our framework takes, as input, a budget which dictates the number of real samples and dependence estimates we are allowed to send each stream window. This formulation allows for heterogenous costs associated with each stream, which accounts for the traffic cost over the WAN. Therefore,}
Constraint \ref{eq:con5} requires that the number of real samples and compact models forwarded across all streams stays within our bound on WAN traffic\footnote{In our implementation, we define our cost function to simply be the amount of data sent over the network. See appendix \ref{appendix:cost} for an evaluation of how our framework handles heterogeneous stream sampling costs.}.
This constraint could be swapped with the objective function to produce a related optimization problem where we directly minimize network traffic subject to a constraint on the accuracy.

Our final constraint (\ref{eq:con6}) simultaneously addresses two related modeling concerns:
\begin{enumerate}[leftmargin=*,noitemsep]
    \item \textit{Underestimating the variance}. Our models estimate the expected value of a target stream given a predictor stream. Therefore, each time we impute a value we are implicitly reducing the variability of the target stream sample and biasing the result of a \textrm{VAR} or \textrm{MAX} query. If the streams are not sufficiently correlated, we could severely underestimate the variability in the target stream.

    \item \textit{Model quality}. If we construct a poor model for the expected value, our imputed values will not be representative of the target stream. Therefore, we require a mechanism for controlling model quality.
\end{enumerate}
\noindent
To address both concerns, consider the following variance decomposition formula:

\begin{equation}
    \Var[X_i] = \E[\Var[X_i|X_{p_i}]] + \Var[\E[X_i|X_{p_i}]]
\end{equation}
\noindent
The $\Var[\E[X_i|X_{p_i}]]$ term captures the variance in $X_i$ that is explained by our model for the expected value.
Since $\Var[\E[X_i|X_{p_i}]] \leq \sigma_i^2$, we introduce a constraint that allows the user to dictate how large of a reduction in variance they can tolerate. This is achieved by computing an expression for the bias introduced by our imputation and bounding it above by a constant ($\epsilon_i$).
A query for the variance executed at the cloud computes the following:
\begin{align}
    \hat{\sigma}_i^2
    &= \frac{1}{\nri + \nsi-1} \sum_{j=1}^{\nri + \nsi} (x_{ij} - \hat{\mu}_{i})^2 \\
    &= \frac{1}{\nri + \nsi -1} \left( (\nri-1)\hat{\sigma}_{r,i}^2 + (\nsi-1) \hat{\sigma}_{s,i}^2 \right)
\end{align}

\noindent
Here, $\hat{\sigma}^2_{r,i}$ and $\hat{\sigma}^2_{s,i}$ are estimators for the variance of the real and imputed samples respectively from stream $i$. Therefore, 
the expected reduction in variance associated with imputing $\nsi$ values rather than sending them is equal to the bias of our variance estimator ($\hat{\sigma}_i^2$), given by:
\begin{align}
    \textrm{Bias}(\hat{\sigma}_i^2)
    &= \E \left[ \frac{(\nri - 1)\hat{\sigma}^2_{r,i} + (\nsi - 1)\hat{\sigma}^2_{s,i}}{\nri + \nsi -1} \right] - \sigma_i^2 \\[1em]
    &= \frac{(\nsi - 1) \Var[\E[X_i|X_{p_i}]] - \nsi \sigma^2_{i}}{\nri + \nsi -1}
    \label{eq:bias}
\end{align}

\noindent
Using this formula, we can readily compute how much we expect the variance to decrease given $\nri$ and $\nsi$.
In practice, the bias will heavily depend on the type of imputation being performed. If we simply replace missing values with a point estimate for the mean, then $\Var[\E[X_i|X_{p_i}]]$ is exactly zero. Model-based imputation techniques will almost always explain a positive amount of the variance. 

\subsubsection{Optimization at the Edge}
Our optimization problem has a non-linear objective function and integer-valued variables, making it a non-linear integer program, which is computationally NP-hard to solve in general. Our system solves this problem at the edge after each tumbling window; therefore, computational efficiency is important. To simplify the problem, we propose treating $p$ as a constant, i.e., we treat the optimal selection of stream predictors as a sub-problem, which we solve prior to solving the optimization problem. We propose a heuristic in section \ref{sec:predictors} for obtaining a solution to this sub-problem efficiently, which leads to the following result.

\noindent
\textbf{Theorem. } \textit{The problem in \probl\;is a convex optimization problem when $p$ is treated as a constant and the integer constraints are relaxed on the objective variables.} \\[10pt]

\noindent
\textbf{Proof.} Concatenate $\nr$ and $\ns$ to form a single vector $n = (\nr, \ns) \in \R^{2k}_{\geq 0}$. To simplify notation for the Hessian, for $i \in 1\;...\;k$, we define the non-negative variables $\psi_i = 2 w_i^2 \sigma_{i}^2 / (n_i + n_{i+k})^3$. Then the non-zero Hessian entries are:

\begin{equation*}
     \frac{\partial^2 f}{\partial n_j \partial n_i} =
     \begin{cases} 
     \; \; \psi_i & 1 \leq i \leq k \textrm{  and  } j=i \\[0.75em]
     \; \; \psi_{i-k} & k \leq i \leq 2k  \textrm{  and  } j=i \\[0.75em]
     \; \; \psi_i & 1 \leq i \leq k  \textrm{  and  } j=i+k \\[0.75em]
     \; \; \psi_{i-k} & k \leq i \leq 2k   \textrm{  and  } j=i-k\\[0.75em]
   \end{cases}
\end{equation*}

\noindent
For arbitrary $z \in \R^{2k}$, we have:

\begin{align*}
    z^T(\nabla^2 f(n))z &= \sum_{i=1}^k \sum_{j=1}^k z_iz_j (\nabla^2 f(n))_{ij} \\
     &= \sum_{i=1}^{k} \psi_i(z_i^2 + z_{i+k}^2) + \sum_{i=1}^k 2 \psi_i z_iz_{i+k} \\
     &= \sum_{i=1}^{k} \psi_i(z_i + z_{i+k})^2 \; \geq 0 \\ 
\end{align*}

\noindent
Therefore, the Hessian is positive semi-definite and the objective function is convex. Given the properties of our constraints, the problem is a convex optimization problem (see appendix \ref{appendix:convexity} for additional details).
This convexity result gives us some assurance that the optimization problem can be solved efficiently, which is important given the heterogeneous compute resources available at the edge. Our sampling algorithm at the edge proceeds according to the flow outlined in Algorithm \ref{alg-sampling}.

\begin{algorithm}
\SetAlgoLined
\SetKwInOut{Input}{Input}
\SetKwInOut{Output}{Output}
\Input{$\;$ Window Duration, $t$; Cost bound, $C$, Variation Reduction Bounds, $\epsilon_i$}
\Output{$\;$ Samples and models to send to the cloud}

    Start Timer for $t$ seconds
    
    \While{Aggregation timer is running}{
    
        Insert inbound samples into cache
        
    }
    
    Estimate $\sigma_i^2$\
    
    Use heuristic to select predictors for each stream
    
    Solve optimization problem in \probs $\textrm{ for } \nri$ and $\nsi$
    
    Forward samples and compact models to cloud
 \caption{Stream Sampling Algorithm}
 \label{alg-sampling}
\end{algorithm}

%% file: Sections/heuristics.tex
There are some barriers to directly applying our optimization framework in real-world environments. We now discuss some of those issues and how to handle them in practice.

\subsection{Predictor Selection}
\label{sec:predictors}
Our optimization problem can be solved efficiently if we treat the predictor indexes as constants. If we wanted a globally optimal solution, it would require considering $O(k!)$ possible permutations for the predictors. We use a polynomial-time heuristic that can be used to obtain a set of predictors more efficiently. For a given stream $X_i$, we simply select, as a predictor, the stream that has the strongest correlation with $X_i$, which requires $O(k^2)$ comparisons. This does not necessarily yield the globally optimal solution; it may be better to choose a slightly sub-optimal predictor for a given stream to reduce the total number of overall predictors, which allows us to send real samples for a small number of predictor streams. We evaluate the accuracy of this heuristic and its impact on latency in sections \ref{sec:eval-heuristic} and \ref{sec:eval-opt-lat}.

\subsection{Modeling and Dependence Estimation}
\label{sec:dependence}
There are many different kinds of predictive models that could be used to impute values from a stream.
We assume that the model estimates the expected value $\E[X_{i} | X_{p_i}]$ and can be represented \textit{compactly}. An appropriate model will also depend on the measure of dependence used by the implementation:

\begin{enumerate}[leftmargin=*,noitemsep]
    \item If we use the \textit{Pearson} correlation coefficient as our measure of dependence, a simple linear model will suffice for estimating the expected value.
    \item If we use the \textit{Spearman} correlation as our measure of dependence, a natural model choice could be polynomial regression. Modeling the response stream based on a $3^{\textrm{rd}}$ degree polynomial provides compactness and the flexibility to fit a variety of monotonic functions.
\end{enumerate}

\noindent
Our framework allows for arbitrarily complex models, but our evaluation shows empirically that these simple models can provide substantial benefit in practice. Furthermore, increasing model complexity will increase traffic over the WAN to send model parameters, require more computation at the edge for dependence estimation and model fitting, and increase model inscrutability. Therefore, \textit{simple models are preferable in our proposed system.}

\subsection{Bounding the Bias} 
\label{sec:bias}
The $\epsilon_i$ constants form upper bounds on the amount of bias we are willing to tolerate in our variance estimates for each stream.
While our estimator for the variance is necessarily biased after imputation, that does not imply that, on average, it deviates substantially from the true variance. In fact, we can obtain a closed form expression for a bound on the bias that guarantees we will not perform any worse than a standard sampling method. Unfortunately, including this expression in the optimization causes the problem to be non-convex (see appendix \ref{appendix:mse} for a derivation). We therefore consider alternative methods for computing a bound.
One method for computing $\epsilon_i$ would be to use a percentage of the observed variance in the stream. For example, we could specify $\alpha=0.05$ and set $\epsilon_i = \alpha \sigma_i^2$ which allows us to bias the variance estimate up to 5\% of the observed variance. To avoid the problem of selecting $\alpha$ for each application, we consider a heuristic based on how much uncertainty is in our estimate of $\sigma_i^2$ at the edge. In practice, we actually compute two estimators for the variance in our framework: one at the edge and one in the cloud. Since this edge estimate is necessarily an imperfect representation of what was observed on the device, we can compute the variance of the edge estimator as follows \cite{mood1973introduction}:
\begin{equation}
    \Var[\hat{\sigma}_i^2] = \frac{1}{N_i} \left( \mu_4 - \frac{N_i-3}{N_i-1}\sigma_i^4 \right)
\end{equation}
where $\mu_4$ is the fourth central moment of $X_i$. We consider selecting $\epsilon_i = \sqrt{\Var[\hat{\sigma}_i^2]}$, which requires the bias to fall within one standard error of the unbiased estimate. The intuition is that we allow the amount of bias in the cloud estimator to scale with the amount of uncertainty in our edge estimator; if we have a precise estimate of the variance at the edge, we will force the optimization framework to avoid substantially biasing the estimate computed in the cloud. Similarly, if our estimate at the edge is noisy, we will allow more bias to be introduced in the downstream estimate. We explore the implications of different selection strategies in our evaluation.

\subsection{Independence Assumption}
\label{sec:iid}
Our framework assumes the independence of samples obtained from each stream in a given window. We believe this assumption is sometimes justified in practice, since measurement noise may dominate changes in the signal during a sufficiently small time interval.
However, this is an unreasonable expectation for most time-series applications. In practice, a \textit{thinning} technique \cite{bda} could be used to address this dependence, which is a common method used for obtaining independent samples from a Markov chain.

We could also weaken the independence assumption to allow for $m$-dependence, where data points that are more than $m$ lags apart are assumed to be independent \cite{tseries}. This allows us to add an error term to our objective function which is proportional to the strength of the dependence:
\begin{equation}
    \Var[\hat{\mu}_i] = \sigma^2_i + 2 \sum_{j=1}^{m} \Cov[x_{i,1}, x_{i,1+j}]
\end{equation}
The number of additional terms in this sum is linear in $m$ (and a constant with respect to our optimization objective). Therefore, the convexity of our problem is unaffected.

%% file: Sections/eval.tex
\subsection{Experimental Setup}

\subsubsection{Testbed and Methodology}
We simulate an edge device using a local machine equipped with an Intel i7-7500U processor and constrained to have 1 GB of RAM. Inbound streams are constructed by reading data from files in fixed size windows.
At the end of each window, we call into an optimization module which solves for the real and imputed sample counts. To test the end-to-end system, we stream this data over the WAN using Amazon Kinesis \cite{kinesis}. Our t2-medium EC2 instance in the cloud runs Apache Storm \cite{storm}, which consumes and processes data from Kinesis.

\subsubsection{Datasets}
We consider three real-world datasets for our evaluation: the \textit{Home dataset}, the \textit{Turbine dataset}, and the \textit{Smart City dataset}. The Home dataset contains temperature measurements taken from three homes in Massachusetts \cite{dataset}. The Turbine dataset consists of sensor traces collected from wind turbines in France \cite{turbine-data}. These sensors collect various measurements, including temperatures, power, wind speed, rotor speed, among other things. The Smart City dataset contains measurements across a diverse set of devices located in Aarhus, Denmark \cite{smart_city1}. We aggregated data from several different devices, including:
\begin{itemize}[leftmargin=*,noitemsep]
    \item Weather sensors, which measure a variety of quantities including temperature and humidity.
    \item Pollution sensors, which capture the amount of a chemical compound in the air.
    \item Parking lot sensors, which measure the current capacity of a specific lot.
    \item Traffic sensors, which report counts of cars that travel past a specific location.
\end{itemize}

We prefer the smart city dataset for examining sensitivity, since it is noisy and representative of a real-world scenario.

\subsubsection{System Comparisons}
We consider the following streaming systems in our evaluation:
\begin{itemize}[leftmargin=*,noitemsep]
    \item {\em ApproxIoT:} A stream-sampling system which performs a variant of stratified sampling at the edge~\cite{approxiot}
    \item {\em S-VOILA:} A stream-sampling system which allocates sample sizes based on the observed variability in each stream~\cite{strat_stream}
    \item {\em Mean:} Our proposed system with edge sampling and cloud-side imputation using the mean
    \item {\em Model:} Our proposed system with edge sampling and cloud-side imputation using a compact model
\end{itemize}

\noindent
For each sampling technique, we systematically vary the amount of data we can send to the cloud and report the error rates for each data size. Unless otherwise specified, our proposed method defaults to using the Spearman correlation as our dependence measure and a cubic representation of the conditional expectation. By default, we set the $\epsilon_i$ bounds equal to one standard error of the edge variance estimate, as discussed in section \ref{sec:bias}.

\subsubsection{Queries and Metrics}
We use a variety of aggregate queries to evaluate our framework, including AVG, VAR, MIN, and MAX. As an example, a query for the average value in a window for each device can be written as:
\vspace{0.1cm}
\begin{verbatim}
    SELECT AVG(response), dev_id, win_id
    FROM Response_Table
    WHERE win_id = current_window
    GROUP BY dev_id;
\end{verbatim}
\vspace{0.1cm}

\noindent
If we have temperature data, these queries could be used to obtain the average or max temperature on a given day. In a smart city setting, we might be interested in quantifying the variability associated with traffic patterns. The MIN and MAX queries are important for applications that are interested in outlier detection or top-k results. We deliberately exclude COUNT queries, since our framework can be easily modified to behave similarly to the ApproxIoT system, which can provide exact answers \cite{approxiot}. We emphasize that our framework supports other aggregates and arbitrary selection predicates, since we are performing sampling. We chose the AVG, VAR, and MIN/MAX queries to evaluate our framework's ability to capture the average behavior and the variability present in each stream (including extreme outliers). To quantify the error for each query, we compute the \textit{normalized root mean square error (NRMSE)}, given by:

\begin{equation}
    \textrm{NRMSE}_i = \frac{ \sqrt{ \frac{1}{T} \sum_{j=1}^T (\hat{\theta}_{ij} - \theta_{ij})^2} }{ \bar{\theta}_i }
\end{equation}
\noindent
where $\theta_{ij}$ is the true value of the aggregate for device $i$ in window $j$ and $T$ is the total number of windows in the experiment. We require a normalized RMSE, since each device may have a radically different expected value and variance.

\subsection{Heuristic vs. Optimal Predictor Selection}
\label{sec:eval-heuristic}
We first evaluate our predictor selection heuristic. Since there are a factorial number of predictor combinations, we use only three highly correlated streams from the Home dataset to measure how well our heuristic approximates the global solution.

\begin{figure}
    \centering
    \begin{subfigure}{.48\columnwidth}
      \centering
      \includegraphics[width=.99\linewidth]{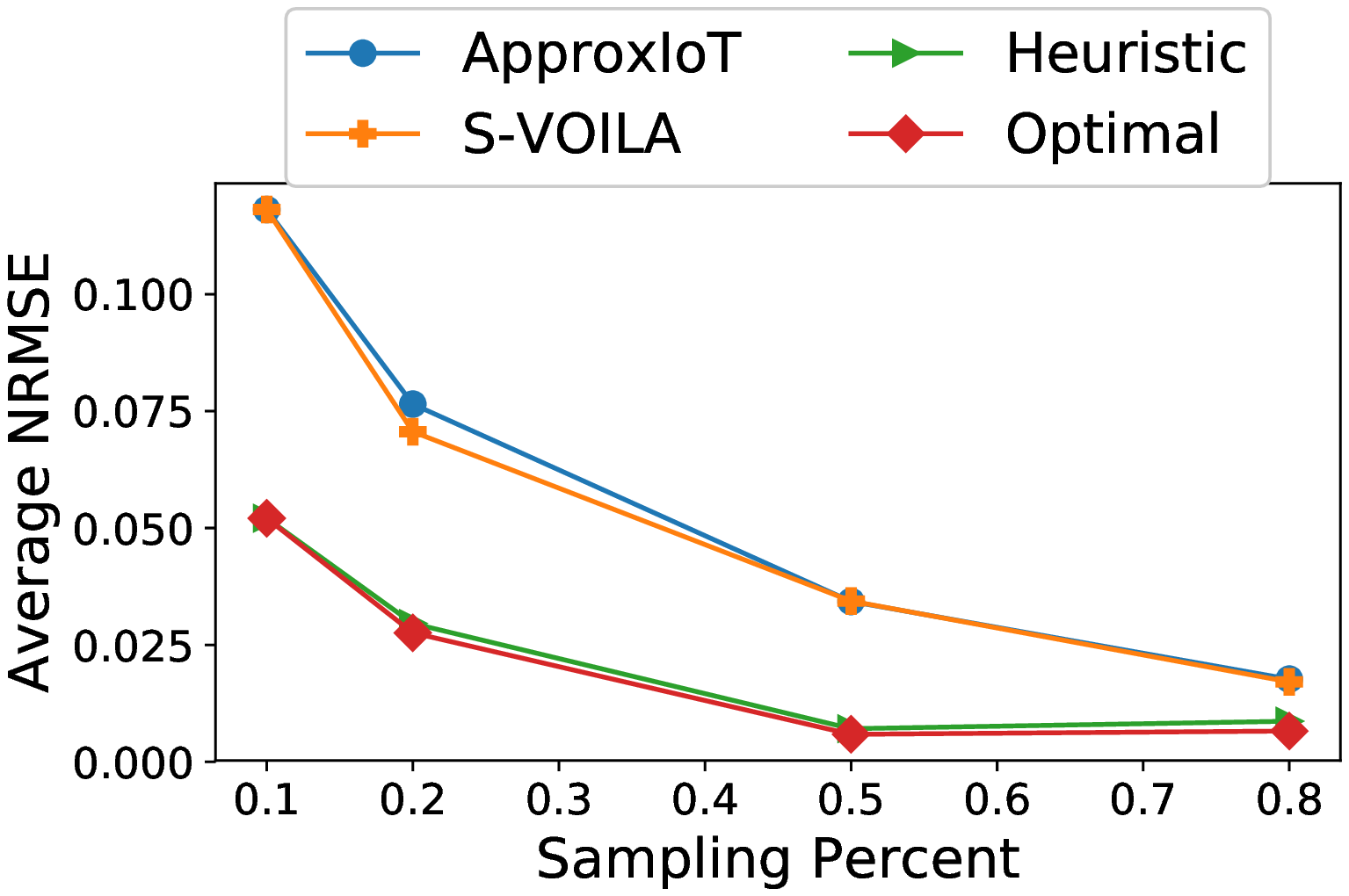}
      \caption{Error rates for an AVG query.}
      \label{fig:glob-sim}
    \end{subfigure}
    \begin{subfigure}{.48\columnwidth}
      \centering
      \footnotesize
      \begin{tabular}{|p{1.2cm}|p{1.0cm}|p{0.8cm}|}
        \toprule
        Method & Average NRMSE & Gain \\
        \midrule
        ApproxIoT & 0.077 & 0.0\% \\
        S-VOILA & 0.071 & 7.8\% \\
        Heuristic & 0.030 & 61.0\% \\
        Optimal & 0.028 & 63.6\% \\
        \bottomrule
      \end{tabular}
      \caption{Performance with a 20\% sampling rate.}
      \label{fig:glob-table}
    \end{subfigure}
    \caption{Heuristic performance on the Home dataset.}
    \label{fig:glob2}
\end{figure}

Figure \ref{fig:glob-sim} compares AVG query errors across different sampling rates. We observe a slight drop in accuracy when using the heuristic compared to the global solution. Figure \ref{fig:glob-table} shows the exact difference in errors for the 20\% sampling rate.
The maximum difference between the heuristic and optimal error reductions across all sampling rates was 3.5\%. 
Our heuristic performs comparably on the other two datasets, with a maximum degradation around 4\%.

\subsection{Turbine Dataset}

\begin{figure*}
    \centering
    \begin{subfigure}{.24\textwidth}
      \centering
      \includegraphics[width=.99\linewidth]{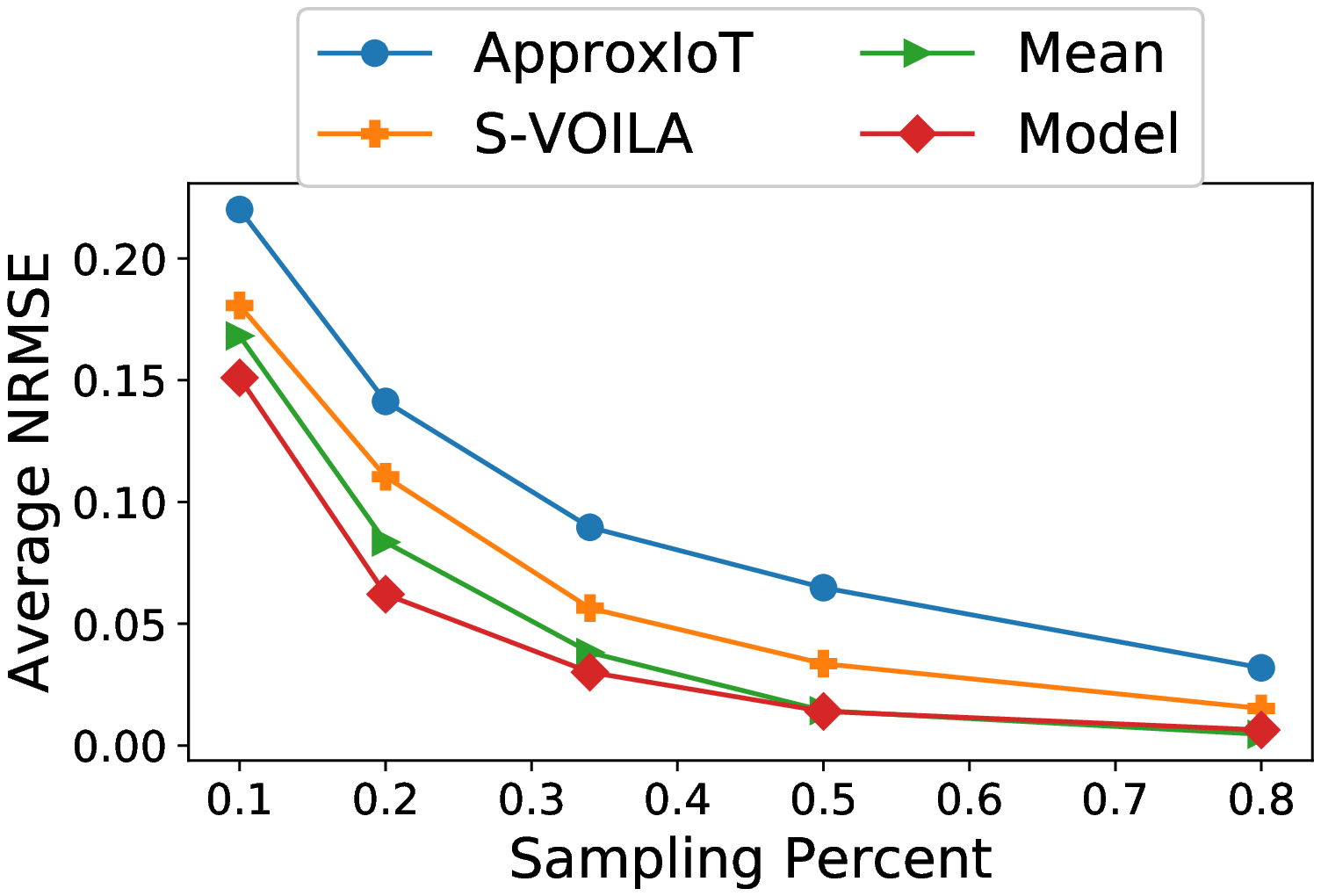}
      \caption{Error for an AVG query}
      \label{mit-avg}
    \end{subfigure}
    \begin{subfigure}{.24\textwidth}
      \centering
      \includegraphics[width=.99\linewidth]{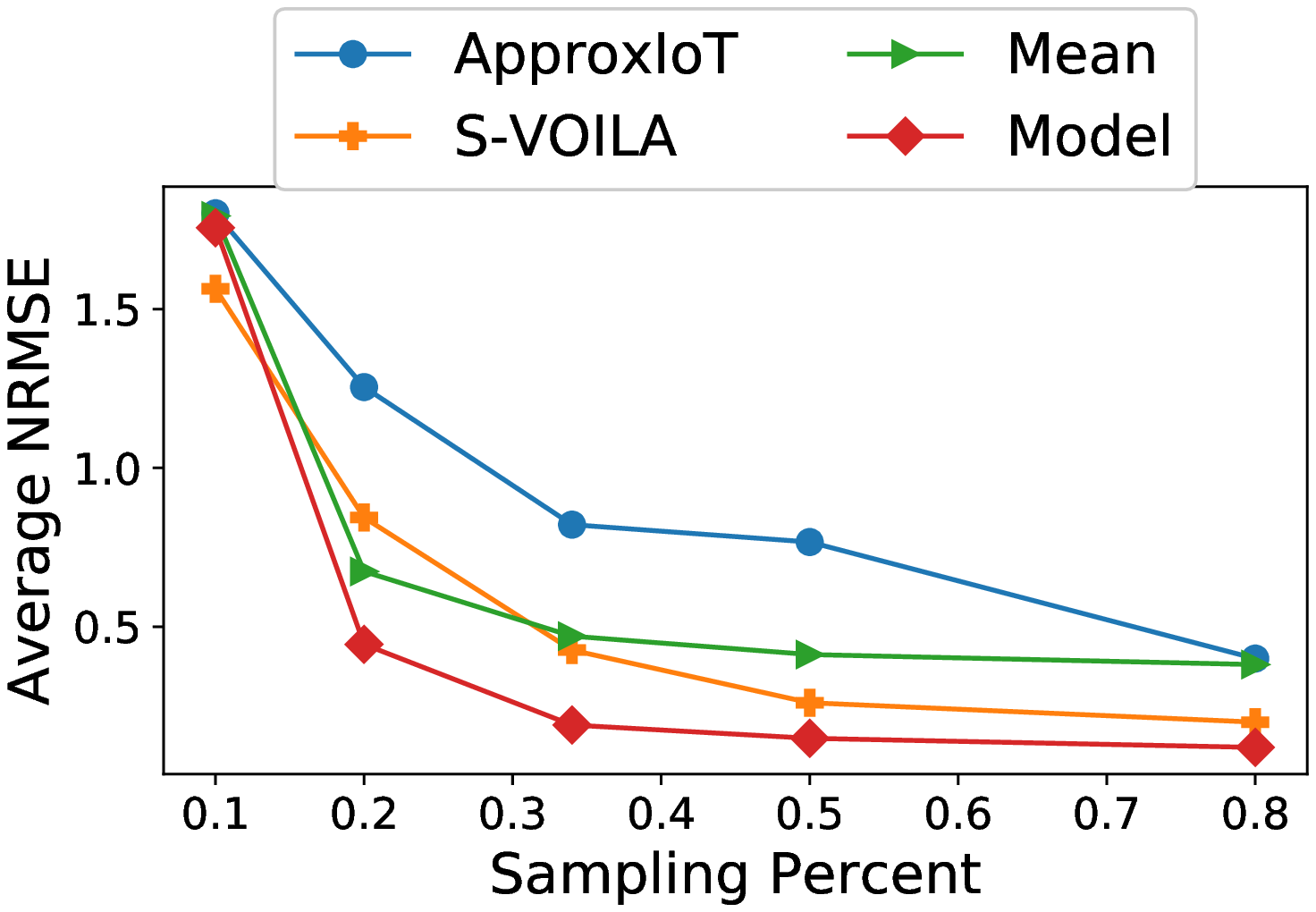}
      \caption{Error for a VAR query}
      \label{mit-var}
    \end{subfigure}
    \begin{subfigure}{.24\textwidth}
      \centering
      \includegraphics[width=.99\linewidth]{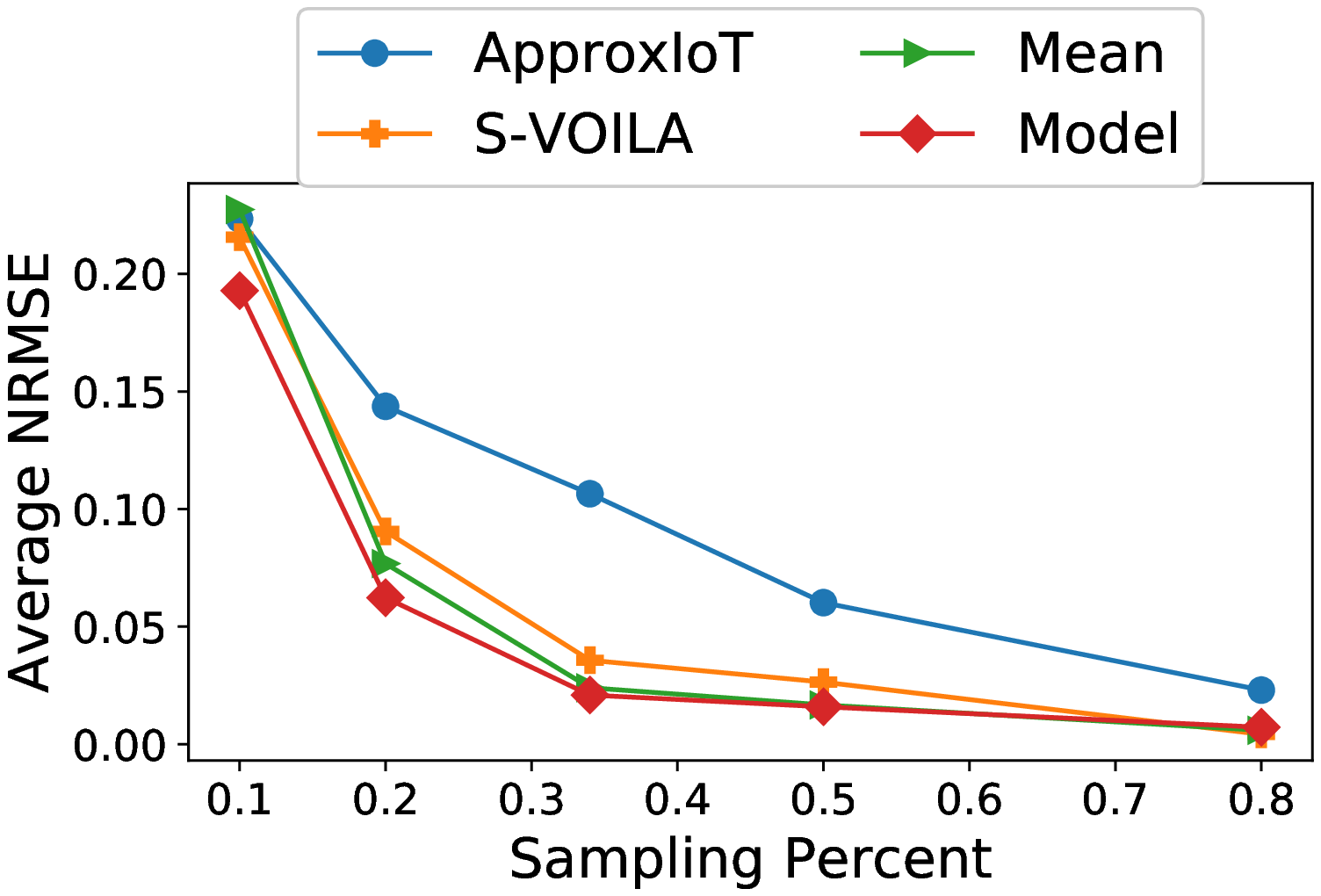}
      \caption{Error for a MIN query}
      \label{mit-min}
    \end{subfigure}
    \begin{subfigure}{.24\textwidth}
        \centering
        \includegraphics[width=.99\linewidth]{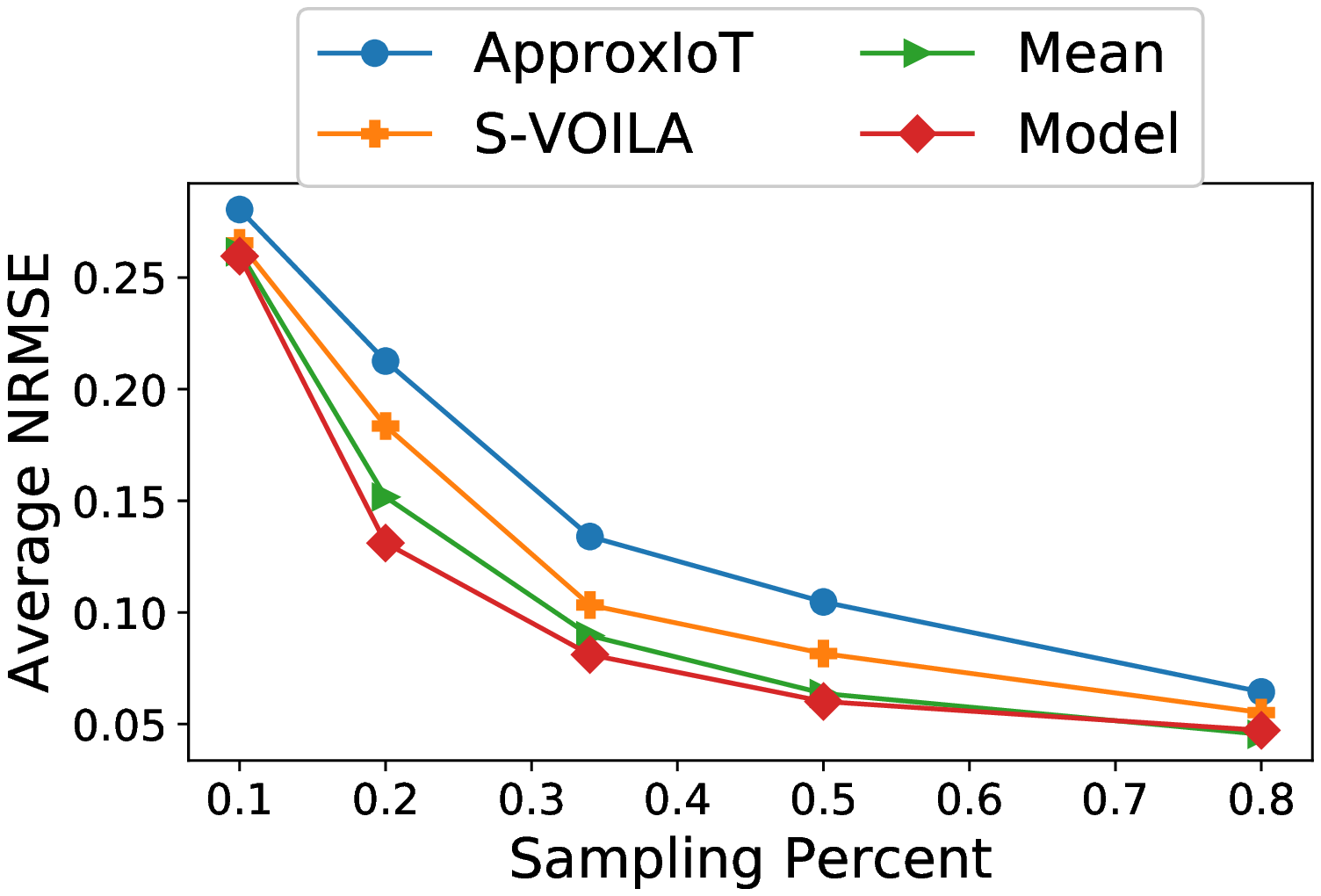}
        \caption{Error for a MAX query}
        \label{mit-max}
      \end{subfigure}
    \caption{Turbine Dataset}
    \label{fig:mit}
\end{figure*}

The Turbine dataset contains measurements from a diverse set of sensors. This application is a perfect fit for our framework, given that bandwidth is often congested in rural locations and low latency is required to react to failures promptly \cite{turbine1}. Since these sensors are in the same location, we readily observe correlation in their measurements and expect our framework to leverage it. The strength of the pairwise linear correlations vary substantially, with some pairs less than 0.05, several pairs in the 0.3-0.5 range, and a few around 0.9.

Figure \ref{fig:mit} shows the performance of our strategies across various sample allowances and aggregate queries. For the AVG query in figure \ref{mit-avg}, we observe that if the user can tolerate an NRMSE of 0.1, our model imputation approach can obtain that level of accuracy sending 16\% of the data, while ApproxIoT requires 32\% of the data, which represents a 50\% decrease. However, this performance varies based on the application's error bound, e.g. obtaining an NRMSE of 0.15 and 0.05 requires 44\% and 60\% less data respectively, when compared to ApproxIoT approach. We observe that S-VOILA consistently outperforms ApproxIoT, since it considers the variance in each stream. Both of our methods also outperform the S-VOILA system, with the model method requiring 27\% and 36\% less data to obtain an NRMSE of 0.1 and 0.05 respectively.

Figures \ref{mit-var}, \ref{mit-min}, and \ref{mit-max} explore how our minimization of the average impacts queries that depend on an accurate representation of the variability. In all cases, we observe an accuracy drop with mean imputation, especially for the VAR query. This is expected, given that imputing values with a constant will necessarily have a greater impact on the variability. The S-VOILA technique is comparable to our model method when at least 80\% of the data is sent over the network. However, on smaller sample sizes our technique outperforms the baseline sampling methods on all four aggregate queries. This effect is caused by the relative importance of the imputed samples. When data is scarce, each imputed sample provides substantial information about the data distribution. This information becomes less important as the number of real samples increases.

\subsection{Smart City Dataset}
For this experiment, we consider devices which have radically different data distributions.
\del{Table \ref{tbl:city} shows a sample of observed absolute linear correlations between different types of devices.}
We observed modest correlations between devices that measure different quantities, e.g., between parking lot occupancy and temperature (0.4 - 0.6). The strength of these correlations vary across stream windows.

\begin{figure*}
    \centering
    \begin{subfigure}{.24\textwidth}
      \centering
      \includegraphics[width=.99\linewidth]{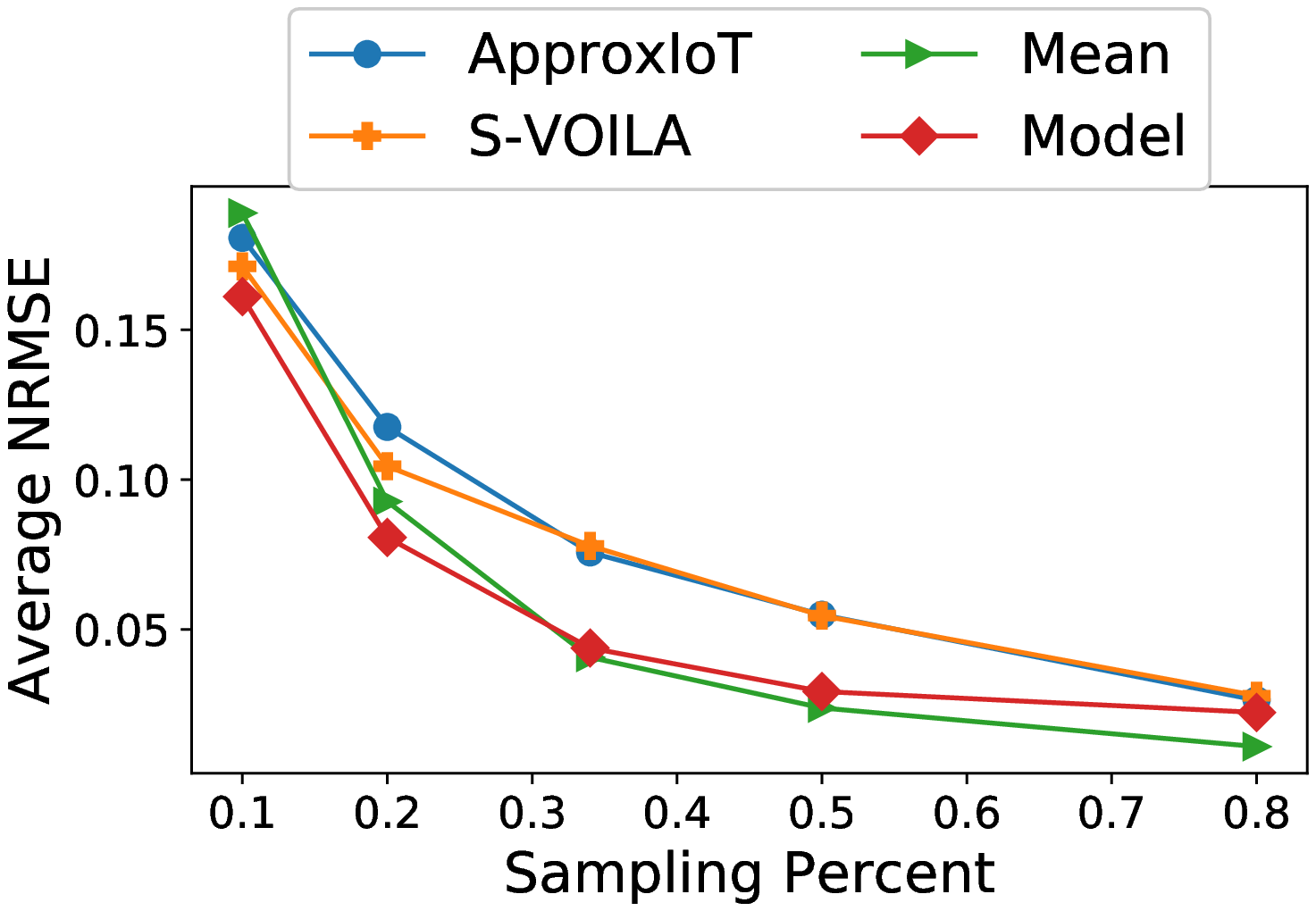}
      \caption{Error for an AVG query}
      \label{city-avg}
    \end{subfigure}
    \begin{subfigure}{.24\textwidth}
      \centering
      \includegraphics[width=.99\linewidth]{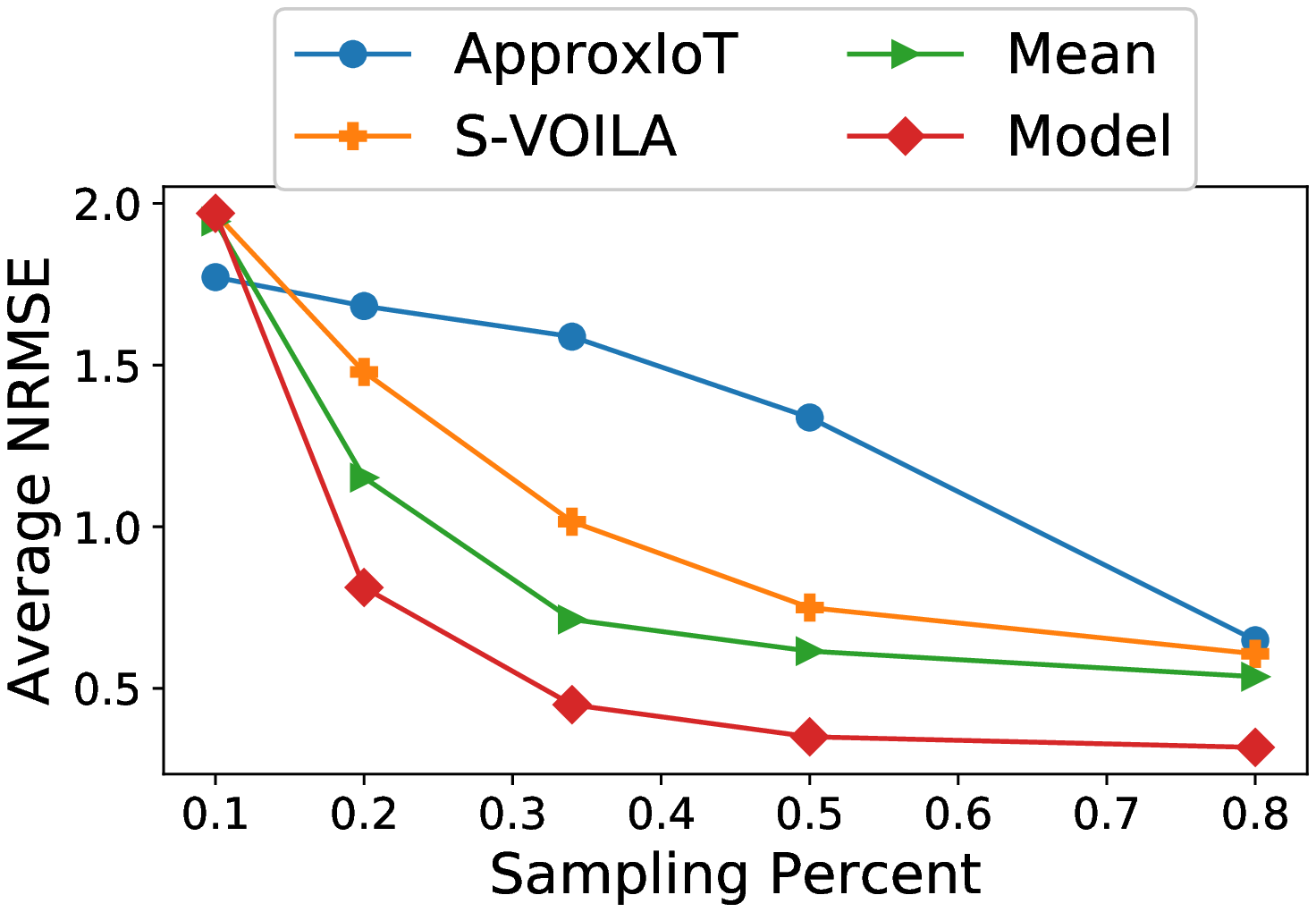}
      \caption{Error for a VAR query}
      \label{city-var}
    \end{subfigure}
    \begin{subfigure}{.24\textwidth}
      \centering
      \includegraphics[width=.99\linewidth]{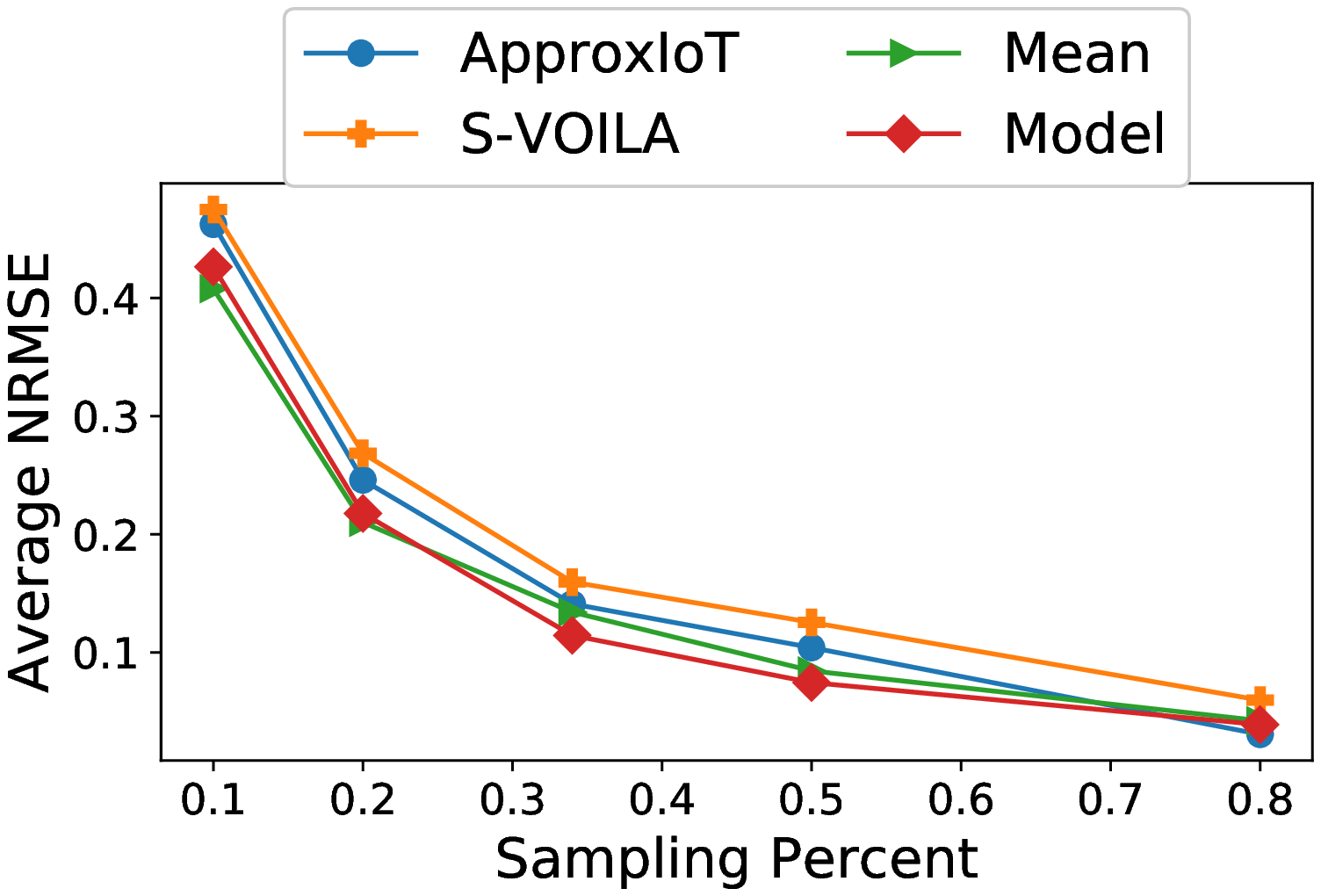}
      \caption{Error for a MIN query}
      \label{city-min}
    \end{subfigure}
    \begin{subfigure}{.24\textwidth}
        \centering
        \includegraphics[width=.99\linewidth]{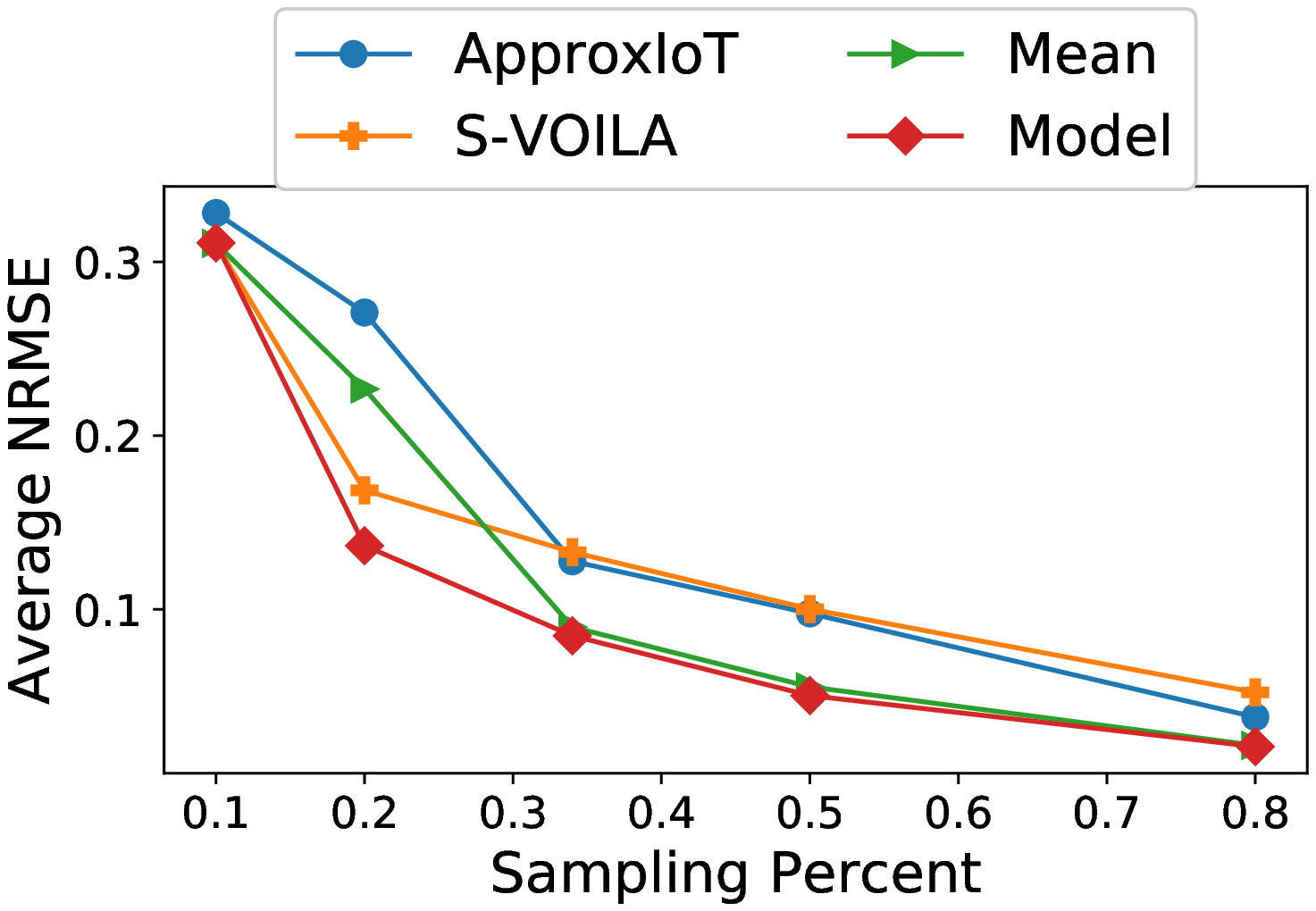}
        \caption{Error for a MAX query}
        \label{city-max}
      \end{subfigure}
    \caption{Smart City Dataset}
    \label{fig:city}
\end{figure*}

Figure \ref{fig:city} shows the performance of our strategies across various sample allowances and aggregate queries. For the AVG query in figure \ref{city-avg}, we observe that if the user can tolerate an NRMSE of 0.1, our model imputation approach obtains that level of accuracy sending 18\% of the data, while ApproxIoT requires 26\% of the data, which represents a 30\% decrease. This performance varies based on the application's error bound, e.g. obtaining a NRMSE of 0.05 requires 42\% less data compared to ApproxIoT. Both of our methods also outperform the S-VOILA system, with the model method requiring 18\% less data to obtain an NRMSE of 0.1 and 42\% less data for an NRMSE of 0.05. We also note that the mean imputation technique begins to outperform the modeling technique once the sample size is sufficiently large. This is expected, since higher sample sizes will allow us to perform more imputation without completely removing the variability from the data. If the number of imputed samples allowed for the model and mean methods are the same, the mean imputation method will necessarily produce a smaller AVG error while inflating the error for the variance. This effect can be observed in figures \ref{city-var}, \ref{city-min}, and \ref{city-max}, where model imputation outperforms the mean approach. We observe that our model imputation approach outperforms the baseline sampling methods on all four aggregate queries.

\subsection{Computational Overhead}
\label{sec:eval-opt-lat}

Our system introduces an overhead after each stream window to model dependence and compute optimal sample sizes at the edge. There is also a computational cost incurred on the cloud side to perform imputation; however, our system uses compact models which are cheap to evaluate. Clouds also have substantial computational resources available to the user, so we exclude it from this analysis.
We used the sequential least squares programming (SLSQP) solver in the Python scipy library to perform the optimization at the edge~\cite{scipy}.

\begin{figure}
    \centering
    \begin{subfigure}{.49\columnwidth}
      \centering
      \includegraphics[width=.99\linewidth]{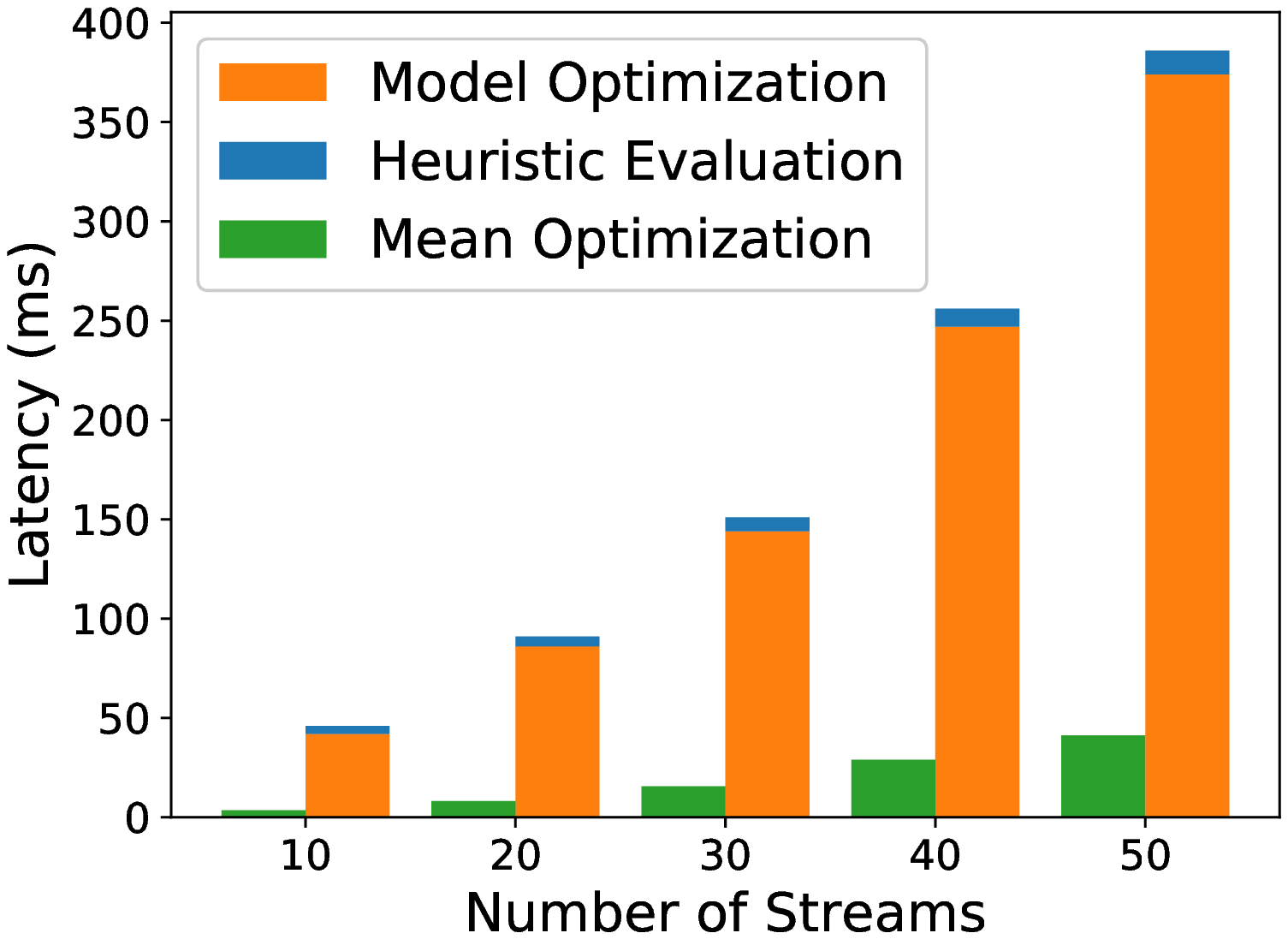}
      \caption{Number of streams vs latency.}
      \label{fig:opt1}
    \end{subfigure}
    \begin{subfigure}{.49\columnwidth}
      \centering
      \includegraphics[width=.99\linewidth]{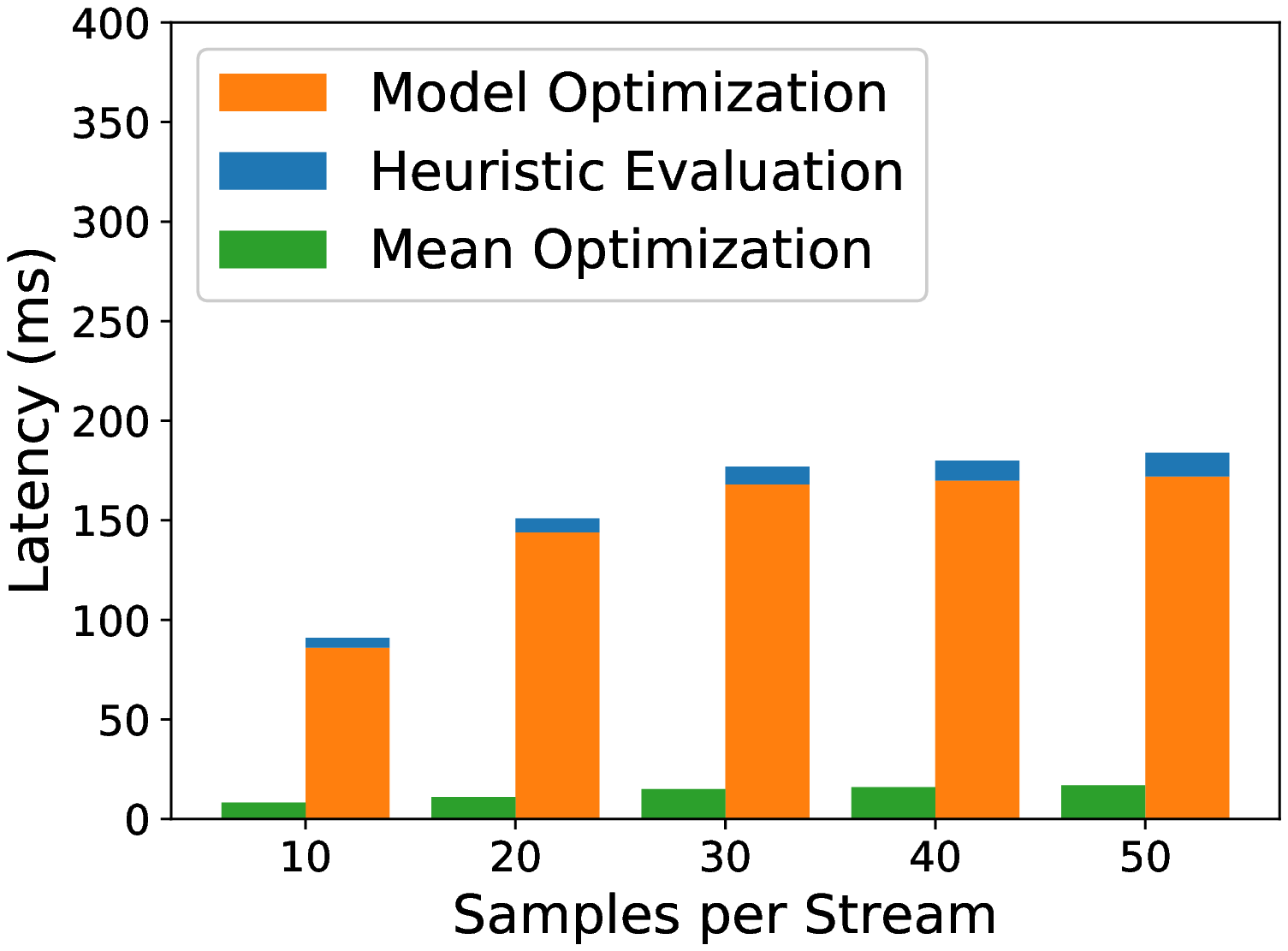}
      \caption{Samples per stream vs latency.}
      \label{fig:opt2}
    \end{subfigure}
    \caption{System Latency}
    \label{fig:opt}
\end{figure}

Figure \ref{fig:opt1} shows how the overall latency scales with an increasing number of streams.
We fixed the arrival frequency at 12 points per stream (e.g. one sample every 5 seconds with a window length of one minute). We observe that the overall latency is less than 400 milliseconds for both imputation methods with a stream count of 50. Solving the optimization problem accounts for the vast majority of the latency compared to the time required to model dependence with our proposed heuristic. We also note that mean imputation requires substantially less time to optimize compared to our model imputation method. There are two reasons for this effect: (1) mean imputation introduces more bias, so we are allowed to impute fewer data points, which restricts the search space in the optimization and (2) almost no time is required to evaluate the heuristic, since the degree of correlation between streams does not impact imputation accuracy. These latencies are unlikely to be an issue if the number of devices is relatively low or the window duration is sufficiently high. Applications that are latency sensitive may need to consider additional parallelism or ensure each edge has a manageable number of client devices.

Figure \ref{fig:opt2} shows the effect of arrival frequency on latency. We observe some effect; however, its impact on performance doesn't appear to grow without bound. We conclude that the dimension of the optimization variable (i.e. the number of streams) has the biggest impact on latency.

\subsection{Sensitivity Analysis}
\label{sec:corr}

\noindent
\textbf{Tolerance for Variance Bias.}
Modeling the expected value biases the variance estimator by reducing the variability present in the sample. We now examine how different tolerances for bias impact error rates using the Smart City dataset and a sampling rate of 50\%. The tolerance values are expressed as a multiple of the standard error (SE) observed in the edge variance estimator, as discussed in section \ref{sec:bias}.

\begin{figure}
    \centering
    \begin{subfigure}{.49\columnwidth}
      \centering
      \includegraphics[width=.99\linewidth]{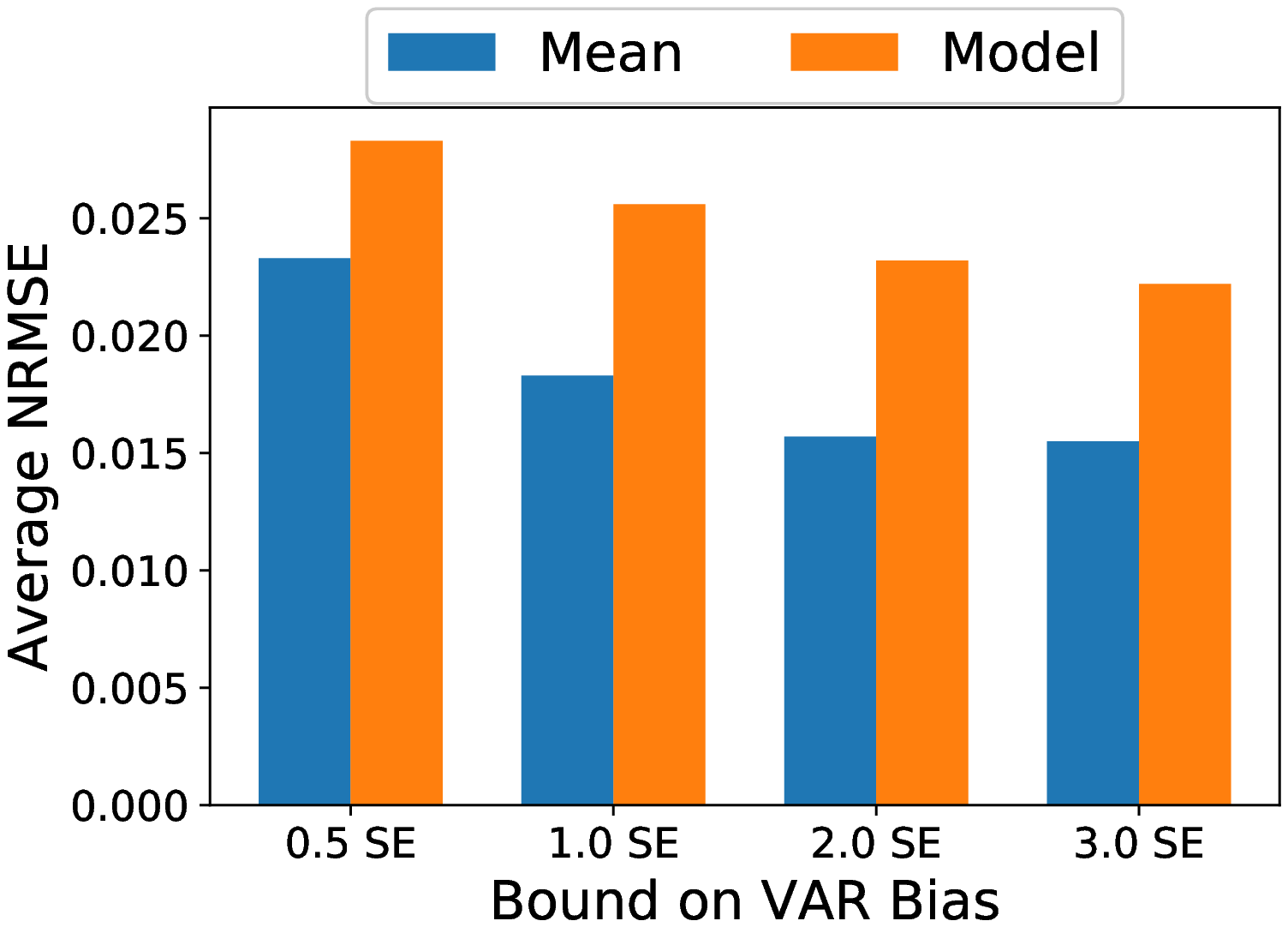}
      \caption{Error for a AVG query.}
      \label{fig:var-bias-avg}
    \end{subfigure}
    \begin{subfigure}{.49\columnwidth}
      \centering
      \includegraphics[width=.99\linewidth]{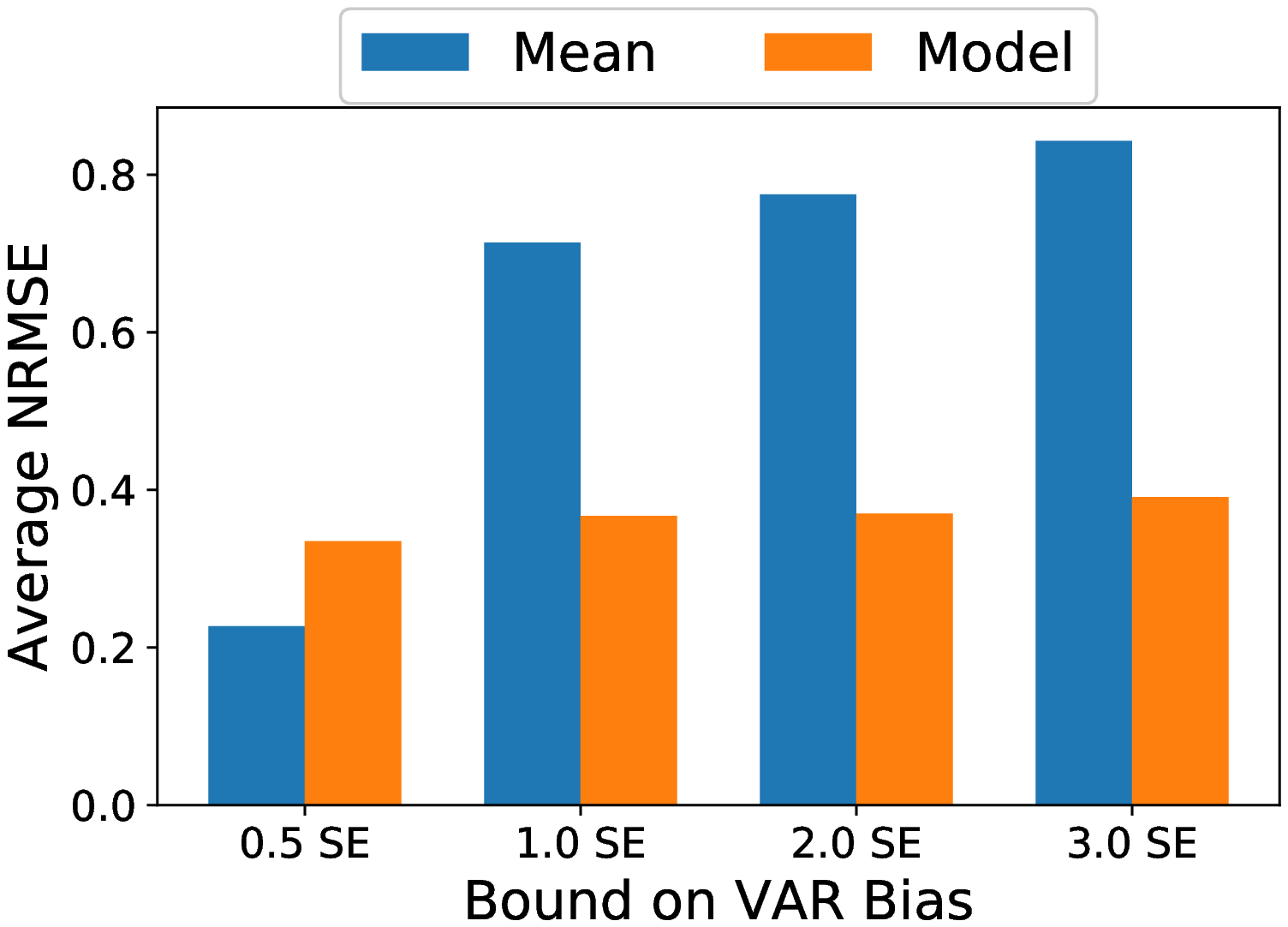}
      \caption{Error for a VAR query.}
      \label{fig:var-bias-var}
    \end{subfigure}
    \caption{Bounding the bias of the variance estimator.}
    \label{fig:var-bias}
\end{figure}

Figure \ref{fig:var-bias-avg} summarizes the error for an AVG query for our mean and model imputation techniques. For both methods, we observe smaller AVG errors as our tolerance for bias increases. This is expected, since a higher tolerance implies we are allowed to perform more imputation with our model for the expected value.

Figure \ref{fig:var-bias-var} shows the impact of these bounds on a VAR query. We observe a steady increase in the error as our ability to bias the result increases. In addition, model imputation method has a smaller impact on the error, since it models the variability more accurately. We also note that mean imputation outperforms model imputation at the 0.5 SE level. This is caused by the difference in imputation frequency between the methods: model imputation is allowed to perform almost 3x as much imputation at that level, since it explains more of the variance. However, this introduces more bias in the variance query. Each application will have a different tolerance for biasing variance queries. Applications mostly interested in the average values will be able to allow higher amounts of imputation compared to those concerned with outliers.

\noindent
\textbf{Correlation Effects.}
The efficacy of our approach depends on our ability to identify correlation in the data and exploit it.
We examine the role of correlation in our framework by generating a synthetic dataset and varying the correlation present in the data. For these experiments, we use two streams drawn from a multivariate normal distribution with means equal to 30 and the diagonal elements of the covariance matrix equal to 16. We systematically vary the off-diagonal elements of the covariance matrix to obtain the desired correlation.

\begin{figure}
    \centering
    \begin{subfigure}{.49\columnwidth}
      \centering
      \includegraphics[width=.99\linewidth]{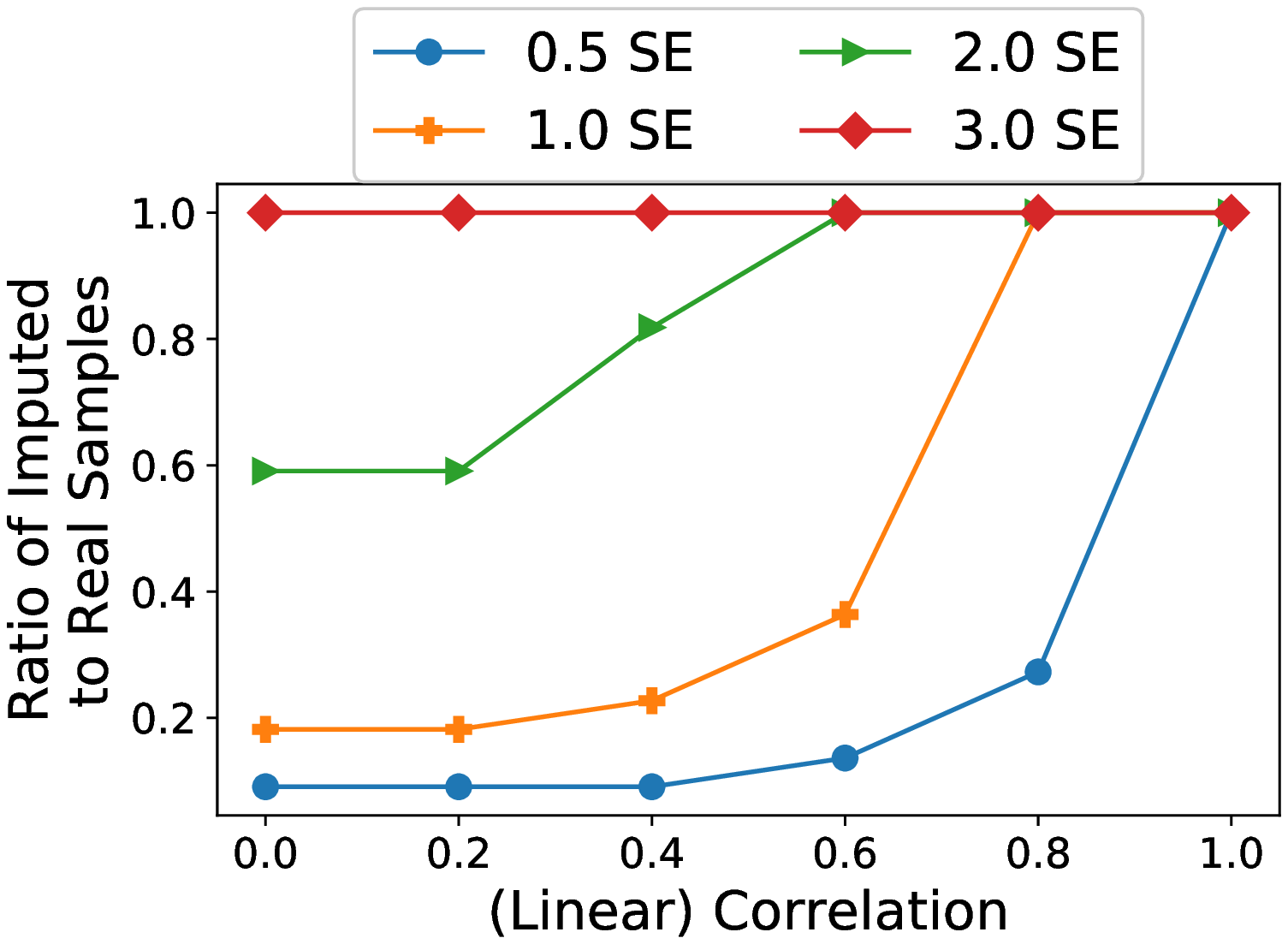}
      \caption{Imputation allowed.}
      \label{fig:corr-sim}
    \end{subfigure}
    \begin{subfigure}{.49\columnwidth}
      \centering
      \includegraphics[width=.99\linewidth]{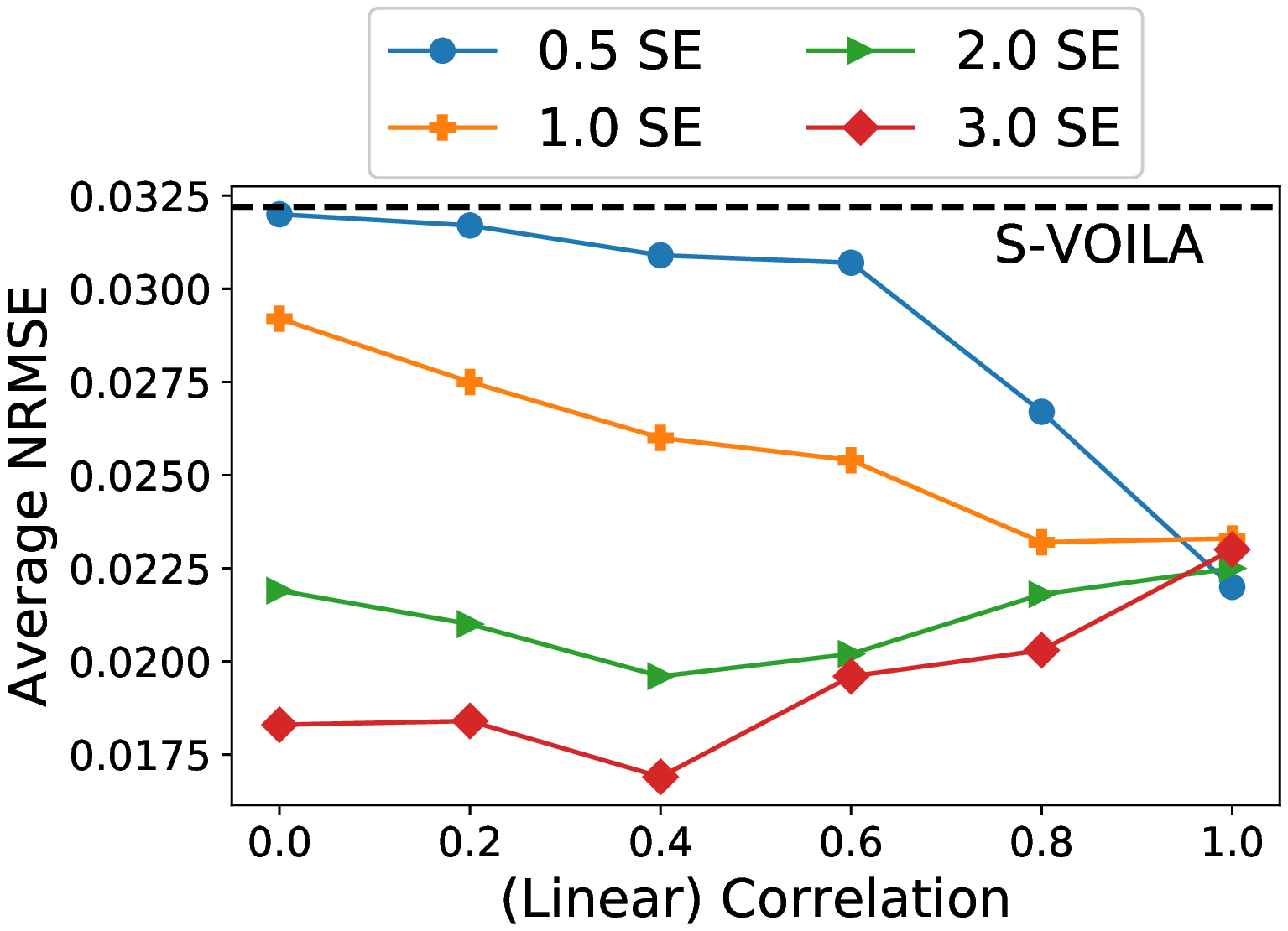}
      \caption{Errors for an AVG query.}
      \label{fig:corr-avg}
    \end{subfigure}
    \caption{Correlation effects on imputation.}
    \label{fig:corr}
\end{figure}

Figure \ref{fig:corr-sim} shows how much imputation we are allowed to perform across multiple values for bias tolerance and correlation.
For 1 standard error, the amount of imputation allowed slowly increases as the strength of the correlation increases. When the correlation is above 0.8, we are allowed to impute a point for every real sample we send over the network.
For 3 standard errors, there is no restriction on imputation, even if the correlation is zero. Using a small tolerance value of 0.5 results in more conservative behavior, requiring very strong dependence before significant imputation is performed.

Figure \ref{fig:corr-avg} shows the AVG query error rates across correlation levels. Our imputation methods always perform as well as the S-VOILA system, regardless of correlation strength.
For smaller bias values (0.5 and 1.0 SE) we obtain a steady decrease in error as the correlation between streams increases. However, for larger bias values (2.0 and 3.0 SE), there is a more complex relationship between correlation and error. The largest errors for these values is obtained when the streams are very strongly correlated. This is explained by considering the difference between mean imputation and model imputation. When the correlation is very low, these techniques are essentially performing mean imputation, since only a small fraction of the variance is explained by the model. As the strength of the correlation increases, the model begins explaining more of the variance, but is a noisier representation of the expected value when compared to mean imputation.

\noindent
\textbf{IID Assumption.}
Time series data often exhibit autocorrelation; however, our framework assumed that samples for each device were IID within a stream window. In section \ref{sec:iid} we discussed options for relaxing this assumption in practice, including implementing a thinning mechanism or adding a penalty term for the autocorrelation. We use the Smart City dataset to evaluate the performance of these options in practice.

\begin{figure}
    \centering
    \begin{subfigure}{.49\columnwidth}
      \centering
      \includegraphics[width=.99\linewidth]{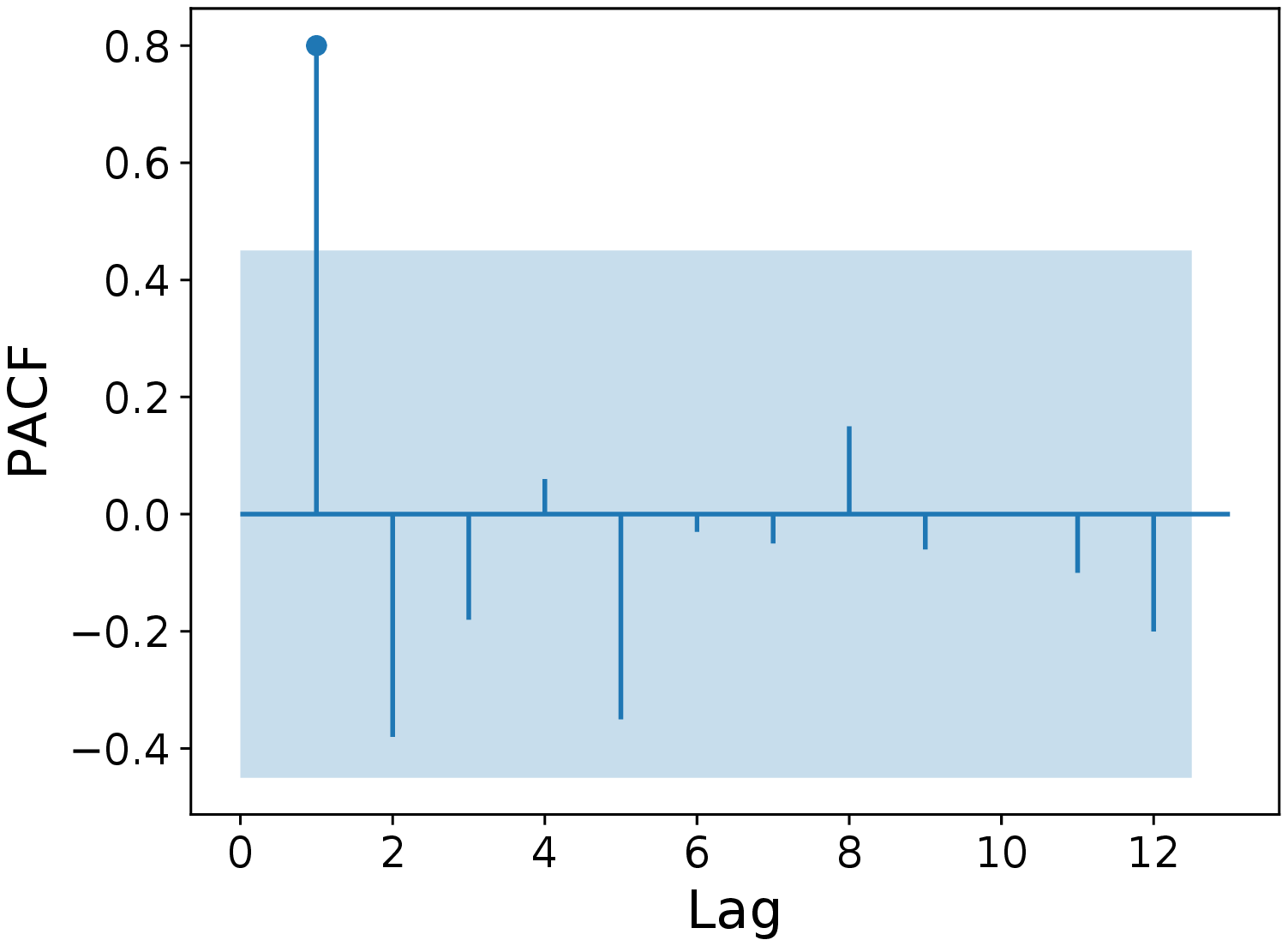}
      \caption{PACF for the $\textrm{NO}_2$ stream.}
      \label{fig:iid-pacf}
    \end{subfigure}
    \begin{subfigure}{.49\columnwidth}
      \centering
      \includegraphics[width=.99\linewidth]{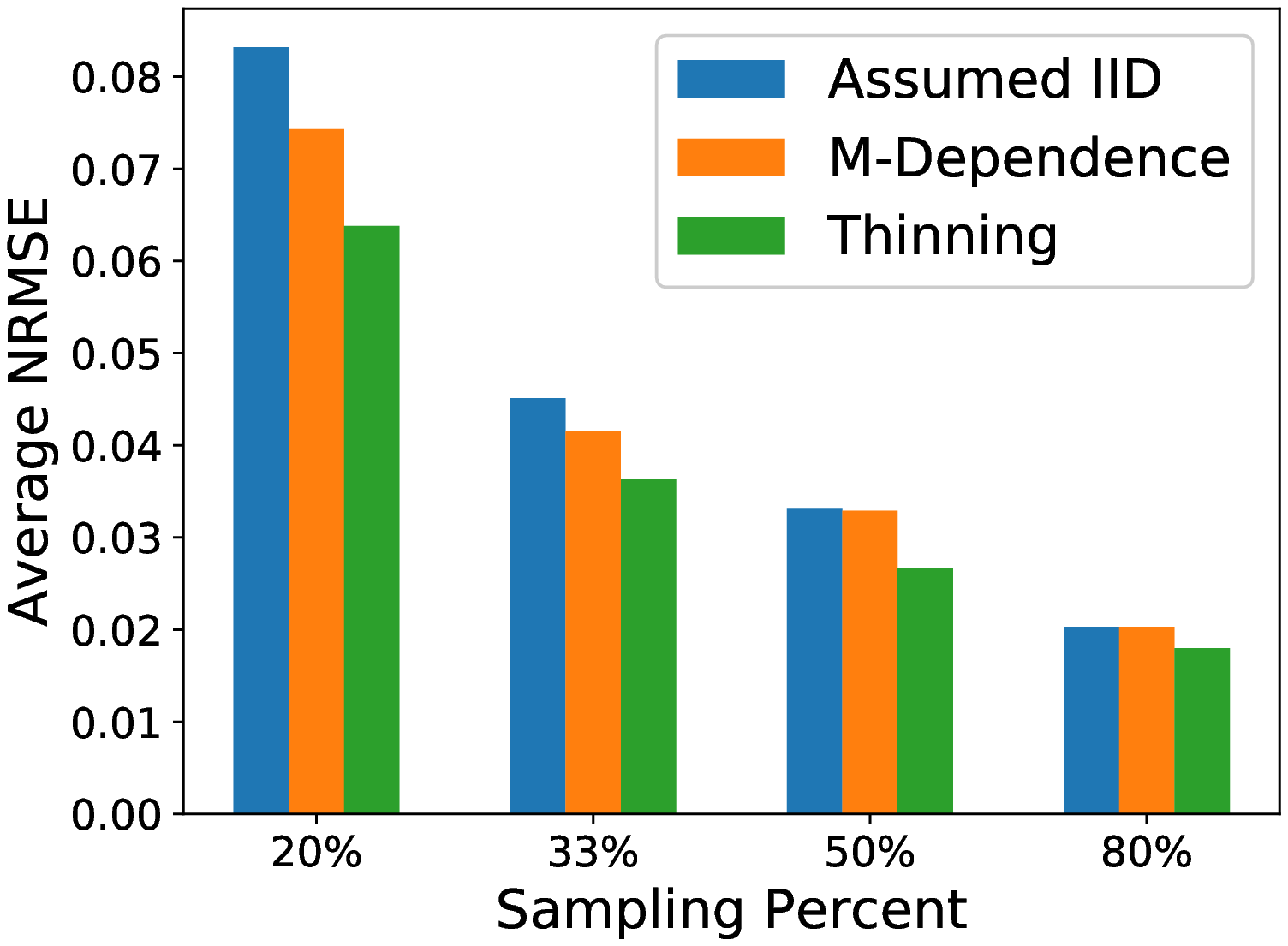}
      \caption{Errors for an AVG query.}
      \label{fig:iid-strategy}
    \end{subfigure}
    \caption{IID experiments Smart City Dataset}
    \label{fig:iid}
\end{figure}

Figure \ref{fig:iid-pacf} displays a partial autocorrelation function (PACF), which shows the strength of the autocorrelation for one of the pollution sensors. There is only one statistically significant lag with a positive correlation around 0.8, so we let $m=1$ for this experiment when estimating $m$-dependence.

Figure \ref{fig:iid-strategy} compares the performance of these techniques on an AVG query. The thinning technique consistently obtains the lowest error rate. Furthermore, thinning works without significant tuning from the user. Implementing $m$-dependence in practice would require dynamically computing a PACF to estimate the number of significant lags in a time series. We therefore recommend the thinning technique in practice.

\begin{figure}
    \centering
    \begin{subfigure}{.49\columnwidth}
      \centering
      \includegraphics[width=.99\linewidth]{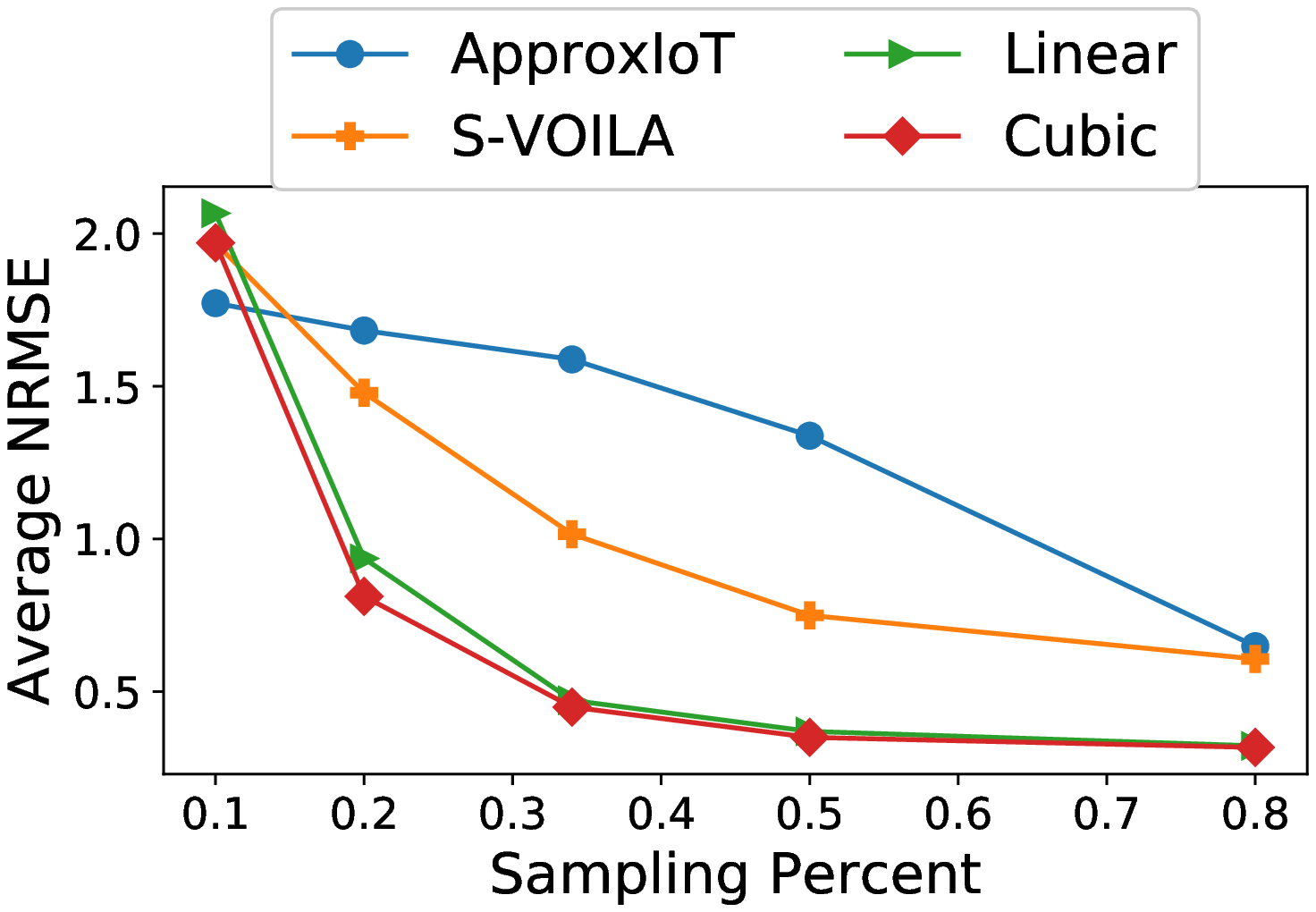}
      \caption{Errors for a VAR query}
      \label{fig:cubic-var}
    \end{subfigure}
    \begin{subfigure}{.49\columnwidth}
      \centering
      \includegraphics[width=.99\linewidth]{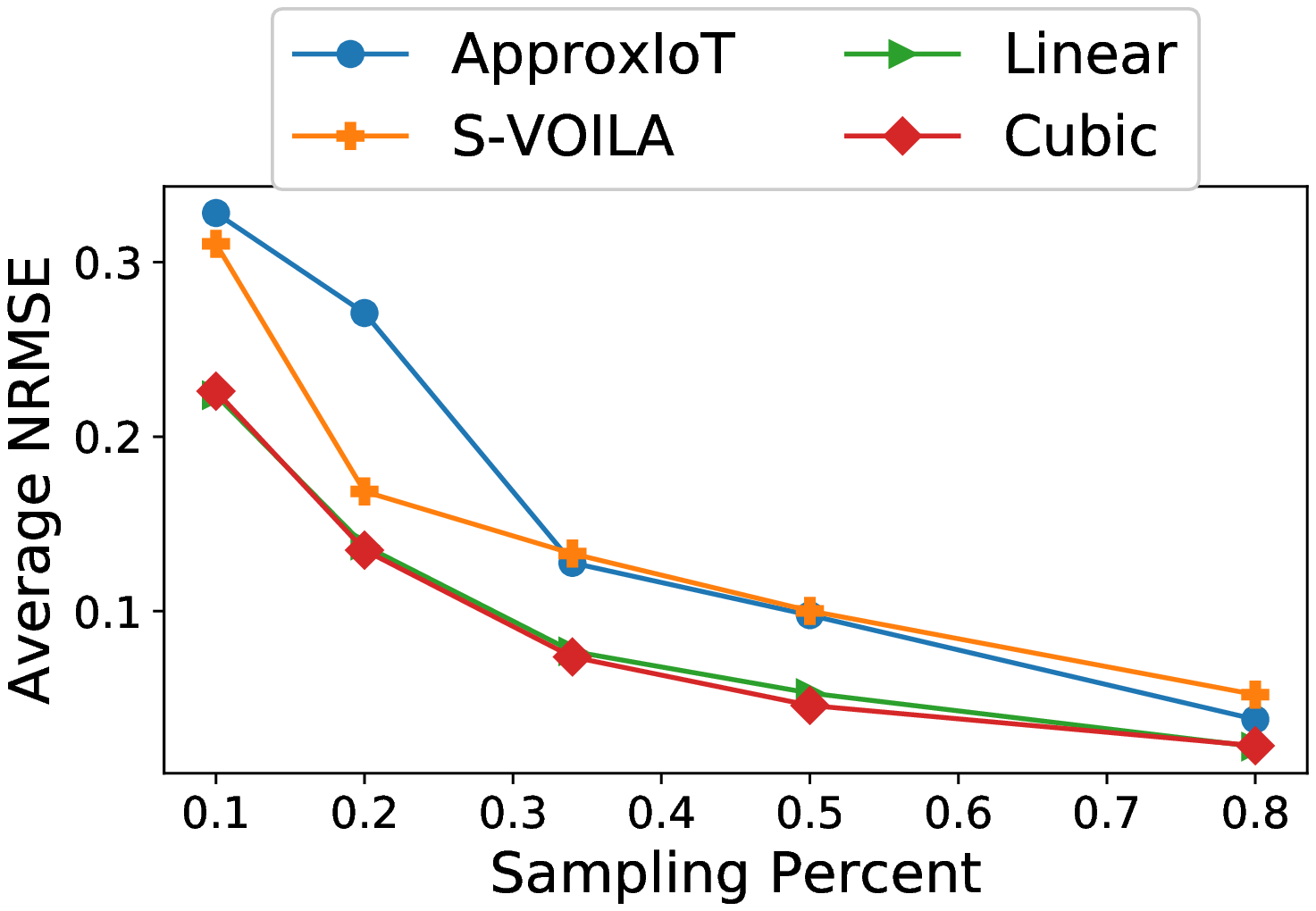}
      \caption{Errors for a MAX query}
      \label{fig:cubic-max}
    \end{subfigure}
    \caption{Linear vs Cubic models with the Smart City Dataset}
    \label{fig:cubic}
\end{figure}

\noindent
\textbf{Linear vs Cubic Models.}
In section \ref{sec:dependence}, we discussed options for measuring dependence between device streams and \textit{compact} representations of the conditional expectation. Our experiments used the cubic representation by default; however, we observe no significant difference between the two methods on an AVG query. The results of the VAR and MAX queries are show in figures \ref{fig:cubic-var} and \ref{fig:cubic-max} respectively. The cubic models provide a slight improvement (around 3\%) when estimating these queries due to their ability to capture the tails of the distributions more accurately. A future enhancement could use LASSO or some other regularization method to dynamically trade-off between linear and cubic models~\cite{lasso}.

\subsection{Discussion}
\label{sec:discussion}
Our sampling and imputation technique generates much less WAN traffic and obtains comparable error rates compared to other stream sampling systems. The bias tolerance $\epsilon_i$ is an important tuning parameter, which bounds the amount of bias in the variance estimator and implicitly dictates the model quality required to perform imputation. The correlation strength also affects our ability to minimize WAN traffic; high correlations between streams provides more opportunity for cost savings.

Our framework is restricted to using a single predictor stream when building models for imputation. It is conceivable that using multiple streams as predictors could produce better models and allow us to impute more values. In this work, we assume model estimation is performed locally at the edge. Leveraging multiple predictor streams increases the complexity of model fitting and increases the size of the model that must be sent over the WAN. In this case, our framework could potentially be modified to leverage historical data in the cloud to identify more complicated relationships between streams. However, our experiments provide some empirical evidence that using a single predictor stream can provide significant cost savings in practice.

%% file: Sections/related.tex
\subsection{Edge and Stream Sampling}
Stream sampling is a common technique for reducing network traffic. Reservoir sampling is frequently used to obtain a variety of probability samples over unbounded streams~\cite{reservoir, heterogeneous1}.
A similar work considers ways in which a central node can obtain a random sample from geo-distributed edge nodes and establishes lower bounds on the number of messages that must be sent~\cite{wood1,wood2}.
A wide variety of other works exist in the field of Approximate Query Processing which seek to systematically trade accuracy for performance \cite{blink,congress}. The ApproxIot and S-VOILA streaming systems both leverage more sophisticated sampling strategies to address properties of individual streams. We improve on these algorithms by supplementing the samples using imputations with explicit bounds on the error.

\subsection{Sensor Networks}
A related area of research considers methods for efficiently obtaining samples from a sensor network. Energy constraints are the main issue as opposed to bandwidth, but the objective is the same: minimize the query error subject to some cost function. The TinyDB system leverages the estimated energy cost and variation for a given device when executing queries~\cite{tinydb}. Dependence is another relevant consideration for this problem, since sensors in the same sensor network will likely produce correlated observations. Systems similar to BBQ are designed with this dependence in mind and attempt to exploit it to reduce the energy requirement~\cite{bbq,sensor2}. We perform an analogous task over dependent data streams by leveraging correlation to reduce the amount of data required to answer queries. Futhermore, we do this in an online fashion (with no offline profiling) and do not assume a parametric model or anything about the underlying data distributions.

\subsection{Sketch Summaries}
Sketching is a widely used technique for computing compact, approximate representations of data \cite{sketching1,sketching2}. In some cases, approximation is not required; for example, moment sketches can exactly estimate the mean and variance in constant space and time \cite{sketching3}. However, our framework makes no assumptions about the downstream operations a user will perform on the resulting data. If a user wants to perform centralized machine learning in the cloud, our framework would provide a dataset which maintains dependencies between streams. Lower fidelity representations, including sketches, would not be as amenable to these use cases.

%% file: Sections/conclusion.tex
Given the increase in data volume and practical WAN bandwidth constraints, efficient data transfer continues to be an important research direction. We present a system that leverages correlation between streams for supplementing samples with accurate imputations in the cloud to reduce WAN traffic. Our evaluation suggests it could reduce bandwidth costs by anywhere from 27\% to 42\%, depending on the application.
While our implementation and evaluation focuses on real-valued data, we believe this general approach can be applied in other streaming settings, such as video streams. Camera deployments often exhibit temporal/spatial correlation and must contend with limited WAN resources~\cite{chameleon}. Given these constraints and additional compute constraints at the edge, it may be preferable to use correlated video streams to obtain approximate query results rather than performing expensive inference at the edge or streaming frames to the cloud for analysis. We leave further exploration of this application to a future work.
Computing and leveraging dependence in these settings could allow for an even broader set of applications to benefit from this approach.

%% file: Sections/appendix.tex
\section{Optimization domain for \probs}
\label{appendix:convexity}
Clearly, $\R^{2k}_{\geq 0}$ is a convex set. To ensure the feasible region is convex, the inequality constraints must be convex functions of the objective variables. The first constraint doesn't apply, since we are treating $p$ as a constant. The next three constraints can be verified by inspection. The last constraint can be rewritten as:
\begin{equation}
\nsi \sigma^2_{i} - (\nsi - 1)\Var[\E[X_i|X_{p_i}]] \leq (\nri + \nsi -1)\epsilon_i
\end{equation}
\noindent
which is an affine function of the objective variable. Therefore, the problem is a convex optimization problem \cite{boyd}.

\section{Exact selection of $\epsilon_i$}
\label{appendix:mse}
The Mean Squared Error (MSE) of an estimator takes into account the amount of bias along with the variation associated with the estimate. If we want to ensure our variance estimate is no worse than a standard technique, we require $MSE_{std}(\hat{\sigma}^2_i) - MSE_{new}(\hat{\sigma}_i^2) \geq 0$. Expanding the expression for the MSE and rearranging, we obtain:
\begin{align*}
    0 \leq& \Var_{std}[\hat{\sigma}_i^2] - \left( \Var_{new}[\hat{\sigma}_i^2] + \textrm{Bias}_{new}(\hat{\sigma}_i^2)^2 \right) \\
    | \; \textrm{Bias}_{new}(\hat{\sigma}_i^2) \; | \leq& \sqrt{\Var_{std}[\hat{\sigma}_i^2] -  \Var_{new}[\hat{\sigma}_i^2]} \\
\end{align*}
\noindent
Then we can rewrite the bound as follows:
\begin{align*} \sqrt{\Var_{std}[\hat{\sigma}_i^2] -  \frac{(\nr - 1)^2\Var[\hat{\sigma}_{r,i}^2] + (\ns - 1)^2\Var[\hat{\sigma}_{s,i}^2]}{(\nr + \ns - 1)^2} }
\end{align*}
which is a non-linear function of the objective variables $\nr$ and $\ns$, violating the requirements for convex optimization. If the dimension of the optimization problem is small, solving it at the edge may be achievable.

\vfill\break
\section{Heterogeneous Sampling Costs}
\label{appendix:cost}
We explore the effect of sampling cost on performance, since our framework can tolerate different sampling costs for each stream. Our baseline sampling strategies are unable to accommodate differing sampling costs, so we compare against the \textit{Optimal Allocation}, which is a version of the Neyman Allocation which includes a unique cost associated with each stream. For these experiments, we used the Smart City dataset and fixed the sampling budget to be one half of the observed data points. We assigned a quasi-random cost to each stream based on a cost distribution, where each experiment specifies a mean and variance.

\begin{figure}
    \centering
    \begin{subfigure}{.49\columnwidth}
      \centering
      \includegraphics[width=.99\linewidth]{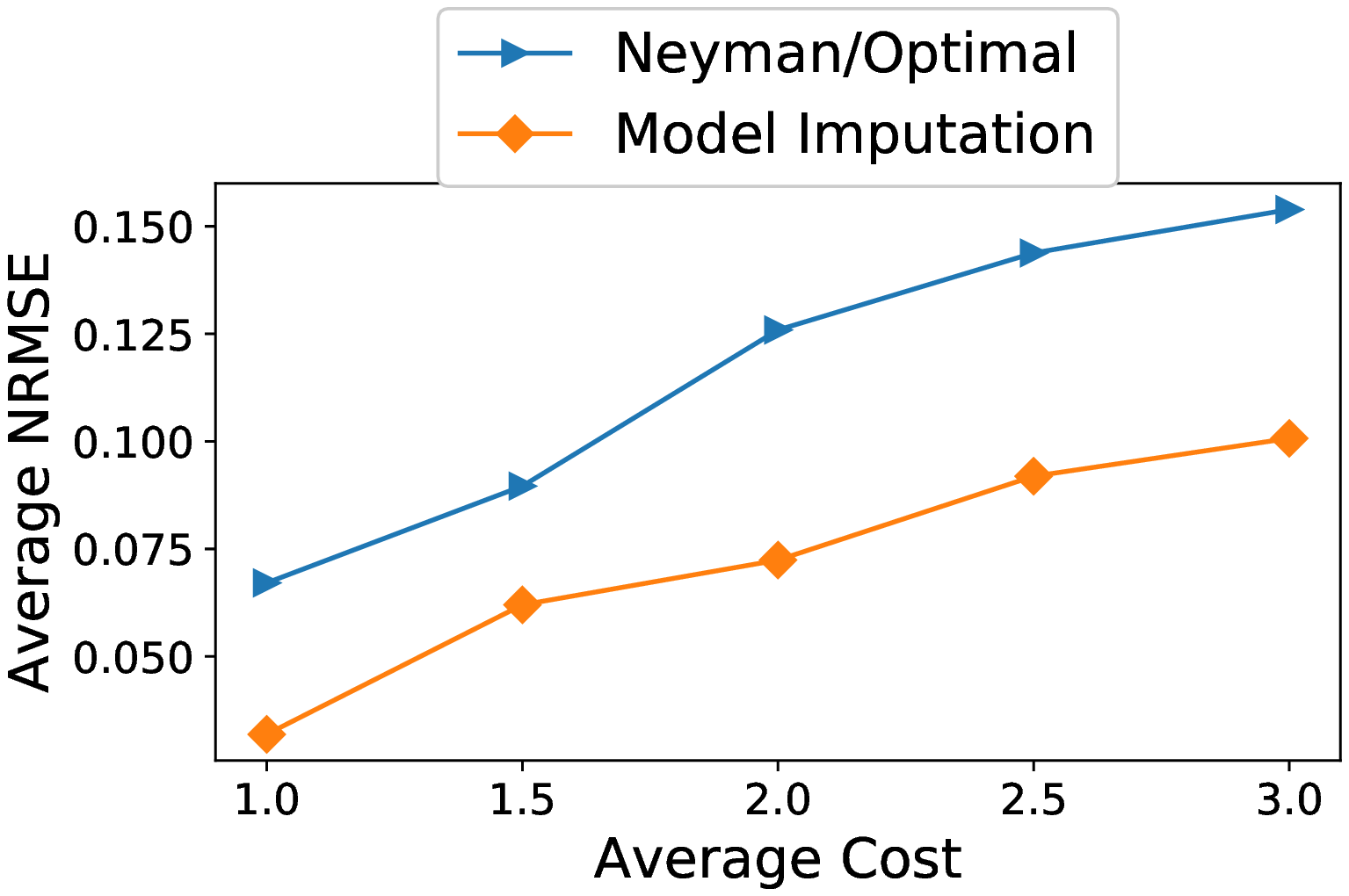}
      \caption{AVG errors with increasing average sampling costs.}
      \label{fig:cost-avg}
    \end{subfigure}
    \begin{subfigure}{.49\columnwidth}
      \centering
      \includegraphics[width=.99\linewidth]{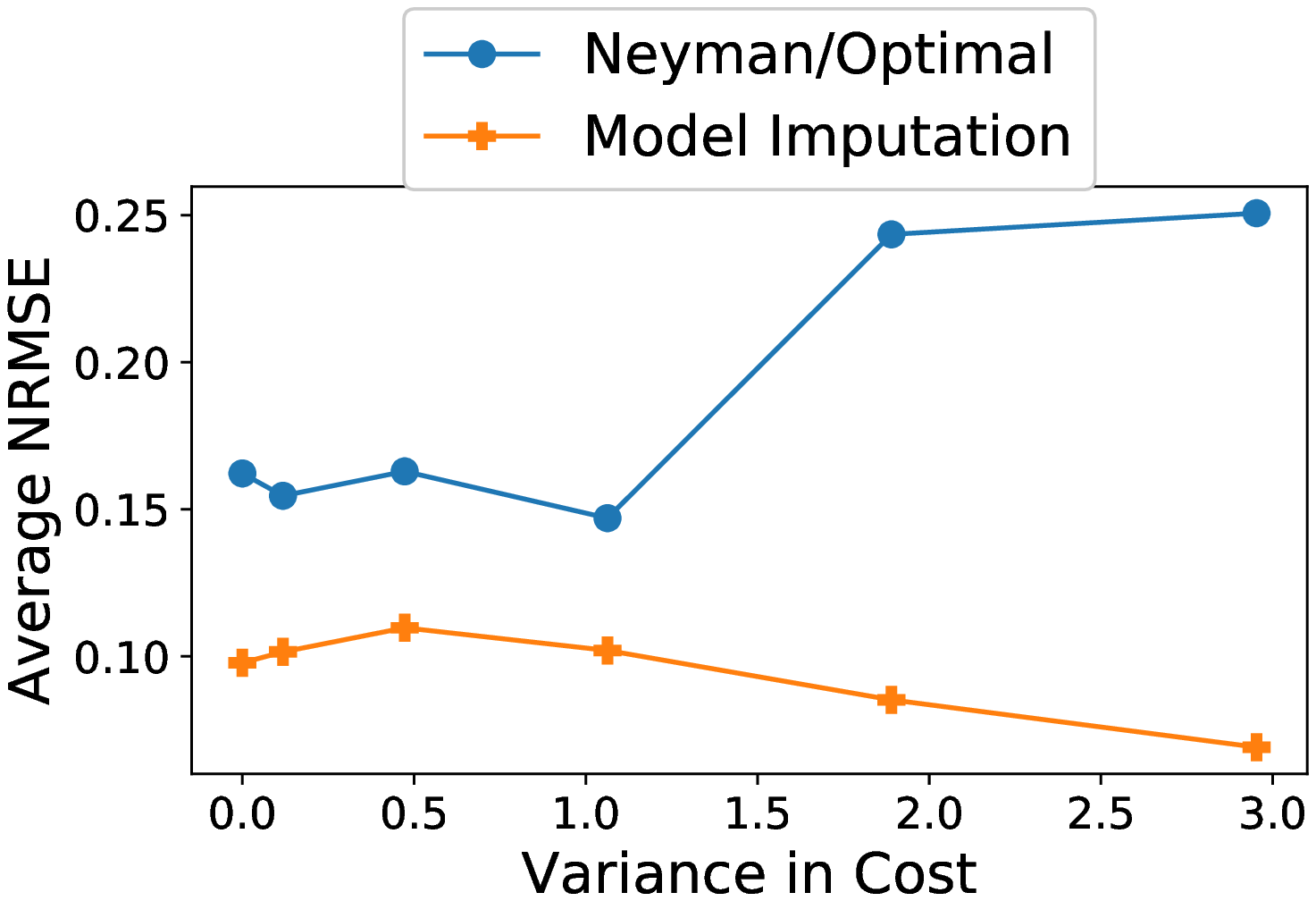}
      \caption{AVG errors with increasing variability in sampling cost.}
      \label{fig:cost-var}
    \end{subfigure}
    \caption{Experiments with heterogeneous sampling costs}
    \label{fig:cost}
\end{figure}

Figure \ref{fig:cost-avg} shows the change in accuracy when we fix the cost variability at 0.25 and slowly increase the average cost of sampling each device. When the average cost per sample increases, the AVG query error necessarily increases for both techniques. We observe that our system assigns more real samples to lower cost devices and imputes values for higher cost devices, resulting in better accuracy. 

Figure \ref{fig:cost-var} shows the results of an experiment where we fix the average sampling cost at 3 and systematically increase the cost variability. We observe that the standard sampling technique effectively leverages the low cost devices when the variance is low, but eventually must continue to sample high-cost devices when the variance increases. Our system is able to consistently improve in accuracy as cheaper devices become available to use for real samples. As long as we have some low cost devices that correlate with the higher cost devices, we are able to send values for the low cost and impute values for the high cost devices.